\documentclass[12pt]{iopart}
\pdfoutput=1
\usepackage[applemac]{inputenc}
\usepackage[english]{babel}
\usepackage[pdftex]{graphicx}
\usepackage{color}
\usepackage{multirow}

\begin{document}

\title[Adiabatic invariants in the Dicke model.] {Adiabatic invariants for  the regular region of the Dicke model.}
\author{M. A. Bastarrachea-Magnani$^1$, A. Rela\~{n}o$^2$, S. Lerma-Hern\'andez$^{1,3}$\footnote{On sabbatical leave from Facultad de F\'isica, Universidad Veracruzana.}, B. L\'opez-del-Carpio$^3$, J. Ch\'avez-Carlos$^1$, and J. G. Hirsch$^1$} 
\address{$^1$ Instituto de Ciencias Nucleares, Universidad Nacional Aut\'onoma de M\'exico, Apdo. Postal 70-543, M\'exico D. F., C.P. 04510}
\address{$^2$ Departamento de F\'{\i}sica Aplicada I and GISC, Universidad Complutense de Madrid, Av. Complutense s/n, 28040 Madrid, Spain} 
\address{$^3$ Facultad  de F\'\i sica, Universidad Veracruzana, Circuito Aguirre Beltr\'an s/n, Xalapa, Veracruz, M\'exico, C.P. 91000}


\begin{abstract}
Adiabatic invariants are introduced and shown to provide an approximate second integral of motion for the non-integrable Dicke model, in the energy region where the system exhibits a regular dynamics. This low-energy region is always present and has been described both in a semiclassical and a full quantum analysis. Its Peres lattices exhibit that many observables vary smoothly with energy, along lines which beg for a formal description. It is shown how the adiabatic invariants provide a rationale to their presence in many cases. They are built employing the Born-Oppenheimer approximation, valid when a fast system is coupled to a much slower one. As the Dicke model has a one bosonic and one fermionic degree of freedom, two versions of the approximation are used, depending on which one is the faster. In both cases a noticeably accord with exact numerical results is obtained. The employment of the adiabatic invariants provides a simple and clear theoretical framework to study the physical phenomenology associated to this energy regime, far beyond the energies where the quadratic approximation can be employed.
\end{abstract}

\vspace{2pc}
\noindent{\it Keywords}: Dicke model, Born-Oppenheimer approximation, integrability.

\maketitle

\section{Introduction}

Twenty years after I. I. Rabi proposed its model for the description of a two-level atom interacting with a single mode of radiation field inside a QED cavity \cite{Rab36}, R. H. Dicke made the generalization for   $\mathcal{N}$ two-level atoms \cite{Dicke54}. Shortly after their proposals, both models were approximated to simpler (and integrable) Hamiltonians by means of the Rotating-Wave Approximation (RWA), i. e. the Jaynes-Cummings (JC), and the Tavis-Cummings (TC) models \cite{JC63,TC68}. The Dicke model has been a topic of discussion for several years, thanks to its most important trait: the prediction of the superradiant phase transition at finite temperature, which in the limit of zero temperature leads to a quantum phase transition (QPT) \cite{Hepp73,Hioe73,Char73,Com74}. However, along many years several authors have pointed out that this phase transition cannot be achieved in a QED cavity systems due to a no-go theorem \cite{Rzaz75}, opening an active discussion, which is far from closed \cite{Knight78,Bia79,Gaw81,Nataf10,Vieh11,Vuk14}. 

The Dicke Hamiltonian is composed of three terms, one describing a monochromatic quantized radiation field inside the cavity,  a second one related with the relative atomic population, and a third which describes the interaction between them. It reads (from now on we set $\hbar=1$)
\begin{equation}
H_{D}=\omega a^{\dagger}a+\omega_{0}J_{z}+\frac{2\gamma}{\sqrt{\mathcal{N}}}\left(a+a^{\dagger}\right)J_{x}.
\end{equation}
Here, the frequency of the radiation mode is $\omega$, associated with the number operator $a^{\dagger}a$. For the atomic part, $\omega_{0}$ is the excitation energy of the single two-level system, while $J_{z}$, $J_{x}$, $J_{y}$ are collective atomic pseudo-spin operators obeying the SU(2) algebra. The $J_{z}$ operator quantifies the relative atomic population. Besides, it holds that if $j(j+1)$ is the eigenvalue of $\mathbf{J}^{2}=J_{x}^{2}+J_{y}^{2}+J_{z}^{2}$, then $j=\mathcal{N}/2$ (the pseudo-spin length) defines the symmetric atomic subspace which includes the ground state. Finally, in the case of cavity QED systems, the interaction parameter $\gamma$ would depend principally in the atomic dipolar moment. 

As mentioned before, the most representative feature of the model is its second-order QPT in the thermodynamic limit, a paradigmatic example of quantum collective behavior. When the atom-field interaction reaches the critical value $\gamma_{c}=\sqrt{\omega\omega_{0}}/2$, its ground state goes from a normal ($\gamma<\gamma_{c}$), with no photons and no excited atoms, to a superradiant phase ($\gamma>\gamma_{c}$), where the number of photons and excited atoms becomes comparable to the total number of atoms in the system, i.e. a macroscopic population of the upper atomic level. 

Being the simplest non-integrable atom-field system exhibiting quantum chaos, the Dicke model became a paradigm in quantum optics, first, and  later in the quantum information community, as it describes more generally the interaction of a set of $\mathcal{N}$ two-level systems (qubits) interacting with a bosonic field, like quantum dots, Bose-Einstein condensates, and polaritons to circuit QED systems \cite{Sch03,Sch07,Blais04,Fink09,Gar11}. Thanks to the advance in sophisticated experimental techniques to control quantum systems, the superradiant QPT was observed experimentally in several systems during the last years. For the Dicke model it was first simulated by means of a Bose-Einstein condensate in an optical cavity \cite{Bau10,Nagy10}. Also, an open version of the model was realized employing Raman transitions \cite{Baden14}, and dynamical non-equilibrium superradiant phase transition has been also observed \cite{Kiner15}. Moreover, by means of  superconducting QED the Rabi model \cite{Niem10,Forn10} and few-atoms Dicke-like models \cite{Cas10,Mezza14} have been explored in this direction.

More than ten years ago, C. Emary and T. Brandes built approximate solutions of the Dicke model by means of the Holstein-Primakoff realization of the SU(2) algebra in the thermodynamic limit \cite{Bran03}. This approach describes the behavior of the model for the ground- and low energy-states, exhibiting the superradiant QPT easily and showing the presence of chaos in the spectrum of the Hamiltonian. Besides, the truncated  Holstein-Primakoff approach makes it possible to extract the critical exponents for the ground-state energy per particle, the fraction of excited atoms, the number of photons per atom, their fluctuations, and the concurrence \cite{Lam04,Vid06}. 

Enhanced by these experimental and theoretical results, and thanks to its algebraic properties, the Dicke model has become an excellent tool from the theoretical point of view for exploring several features of quantum many-body systems with collective degrees of freedom, e.g., the QPT, the Excited-State Quantum Phase Transitions \cite{Per11,Bran13,Bas14,Kloc16} and their relation with the thermal phase transition \cite{Bas162,Perez16}, the onset of quantum chaos and its correspondence to the classical limit \cite{MAM91,Bak13,Bas142,Bas16}, quantum quenches \cite{Paa09,Lob16} and the problem of equilibrium and thermalization in isolated many-body quantum systems \cite{Alt121,Mata15}, to name a few.

This is the motivation for looking for as many analytical or semi-analytical descriptions of the spectra and observables of the model as it is possible. The truncated Holstein-Primakoff approximation decouples the low energy modes into two independent harmonic oscillators. It is valid when the number of excitations is small compared with the number of atoms, allowing for a description of a few low energy states far from the QPT \cite{Cas111,Hirsch13}. The lack of a general analytical description is rooted in the fact that the Dicke model does not have as many integrals of motion as degrees of freedom, being in this sense non-integrable, unlike the TC model. Numerical calculations provide useful results in truncated subspaces  \cite{Chen08,Liu09,Bas11}. 

On the other hand, it has been recently shown that, thanks to being restricted to the single qubit case, the Rabi model can be considered as integrable \cite{Braak11,Chen12}, though the controversy about this fact remains \cite{Moroz13,Batch15}. A similar method to the one which was used for the Rabi model in order to show its integrability, has been recently applied to the Dicke model with three \cite{Braak13} and two atoms \cite{Duan15,He15}. However, a concise analytic solution of the Dicke model is far to be obtained, and in the general case is quite probably impossible.

In this work we construct second integrals of motion in the Dicke model employing the Born-Oppenheimer approximation (BOA) \cite{Berry93}. They are useful and predictive within the non-chaotic energy regime and within a wide range of values of the external parameters, coupling and frequencies, both in the normal and superradiant phases. The BOA validity relies on the fact that part of the system oscillates faster than the other one, causing an effective decoupling where the slow (adiabatically changing) variables enters in the fast dynamics as simple parameters. As the Dicke model has a one bosonic and one fermionic degree of freedom, two versions of the approximation are used, depending on which one is the faster. Similar approaches have already been reported in the Rabi model, the Dicke model, as well as for the JC and TC modes. In the Rabi model the fast atomic BOA has been employed to determine the entanglement of a single atom respect the bosonic field \cite{Lib06}, and to unveil previously unnoticed aspects of the RWA in the JC model \cite{Lar07}. In the Dicke model the fast atomic BOA was applied to study its ground-state properties, like the finite size scaling of the entanglement between the components of the system and other physical observables \cite{Lib062}. Also, it has been used to study the finite size dependence of the tunneling driven ground-state energy splitting in the superradiant phase \cite{Chen07}. The fast boson BOA has been applied to the Rabi model in \cite{Irish05}, and then extended to develop the so called generalized RWA  \cite{Irish07}.  Also, in Refs.\cite{Keeling10,Itin10} the fast boson BOA has been employed to describe the different phases that can be found in the Dicke model realization in Bose-Einstein condensates in optical cavities \cite{Bau10}.    

In this contribution, we extend the use of the BOA to higher energies in the Dicke model, and show that it can describe the regular, non-chaotic energy regime of the model, which extends  from the ground-state to an upper energy (above or below the ESQPT critical energy), depending strongly on the level splitting and the frequency of the field \cite{Cha16}. In both cases, the fast boson and fast atomic BOA, we explicitly derive the approximate second integral of motion that makes regular the low energy regime of the Dicke model, and show its range of applicability in the model parameter space. The BOA extends the theoretical tools to describe the, so far unexplained, whole non-chaotic or regular energy region of the Dicke model, far beyond the region of validity of the quadratic approximation of the Holstein-Primakoff description. We give stringent numerical evidence showing that the BOA, and the second integrals of motion coming from it, shed new light in the study of the quantum dynamics of the model. This contribution extends the study presented in Ref.\cite{Relano16} by some of us, where the basic idea was introduced and complementary results were given for the fast pseudospin BOA.    

The paper is organized as follows, in  section 2 the fast pseudospin approximation and  the fast boson one are presented, and their low energy frequencies are calculated. These frequencies are compared in section 3 with the exact ones to establish the region of validity of the approximations in the Dicke model parameter space. In section 4 we derive a semiclassical approach for the slow variables, which allows us to obtain analytic results for the expectation values of  observables. These results are extensively compared in section 5 with exact results coming from the numerical diagonalization of the Dicke Hamiltonian. In the region of applicability the numerical results are nicely reproduced by the approximations, giving a simple framework to understand them. In section 6 the adiabatic invariants associated with the two fast-slow approximations are unveiled and numerical tests are presented, showing them as approximated integrals of motion both in the semiclassical and quantum versions of the Dicke model until the appearance of chaos in the energy and parameters space. In this same section, the requantization of the semiclassical slow variable is presented and the resulting spectra are compared with the exact ones which come from numerical diagonalizations, showing again, a remarkable accord. Finally in section 7 we give our conclusions.  

\section{Adiabatic approximation}

The Dicke Hamiltonian commutes with the $\mathbf{J}^{2}$ operator, defining different subspaces with fixed $j$. As mentioned before, the maximum value $j=\mathcal{N}/2$ designates the symmetric subspace where the ground-state lies. Inside this subspace, the Hamiltonian only has two degrees of freedom, one for the collective atomic degree of freedom (pseudospin) and one for the bosons. An effective decoupling between them can occur as a consequence of the different temporal scales in their respective intrinsic dynamics, in two different ways. One when the pseudospin dynamics is much faster than the bosonic one, and the second in the opposite case. In the following two subsections we present the two approaches separately. To derive in a simple way the approximations, we start with
a quantum-classical approach, quantum for the fast variables and classical for the slow ones. In the following sections full quantum and full classical results are presented in the framework provided by the fast-slow approximations. 

\subsection{Fast Pseudospin approximation}

For the sake of clarity, let us consider the Dicke Hamiltonian in terms of classical boson variables, i.e. by substituting the annihilation [creation] operators according to $a\rightarrow (1/\sqrt{2})(q+ip)$ [$a^\dagger\rightarrow (1/\sqrt{2})(q-ip)$]
\begin{equation}
H_{PS}=\frac{\omega}{2}(p^2+q^2)+\omega_o J_z+\frac{2\gamma}{\sqrt{j}}q J_x,
\end{equation} 
If we  {\it freeze} the bosonic slow variables, a simple Hamiltonian for the pseudospin variables is obtained, which  can be easily diagonalized by considering a rotation ($\beta$) around the $y$-axis
\begin{eqnarray}
H_{PS} = \frac{\omega}{2}(p^2+q^2)+\omega_P(q) J_{z'}\mbox{\ \ \ \  \ with} \nonumber \\
J_{z'} \equiv \cos\beta J_z + \sin\beta J_x , \,\, 
\cos\beta = \frac{\omega_o}{\omega_P(q)}, \,\,  
\sin\beta = \frac{2\gamma}{\sqrt{j}}\frac{q}{\omega_P(q)}, \label{eq:beta} \\
\omega_P(q) = \sqrt{\omega_o^2+ \left(\frac{2\gamma}{\sqrt{j}}q\right)^2}. \label{eq:rotation}
\end{eqnarray}
The rotation  angle $\beta$ can be positive or negative depending on the sign of $q$. The previous Hamiltonian describes the precession of the pseudospin around the rotated  $z'$ axis  with  angular velocity $\omega_P(q)$, and can be easily diagonalizad by considering eigenstates of the rotated operator $J_{z'}|j m'\rangle= m' |j m'\rangle$    
\begin{equation}
H_{PS}|j m'\rangle =\left[ \frac{\omega}{2}(p^2+q^2)+\omega_P(q) m'\right]|jm'\rangle,
 \label{Hmp}
\end{equation}
\noindent
with $m'=-j,-j+1,...,j-1,j$.  The  eigenvalues, $H_{m'}(p,q)$, depend on the, up to now, {\it frozen} bosonic variables. If we let these variables evolve, these eigenvalues  define, for each $m'$, an effective Hamiltonian for the slow bosonic variables, which has the standard form of a classical particle moving in a conservative potential 
\begin{equation}
V_{m'}(q)=\frac{\omega}{2}q^2+\sqrt{\omega_o^2+ \left(\frac{2\gamma}{\sqrt{j}}q\right)^2}m'. 
\label{eq:vm}
\end{equation}

We have manipulated the Dicke Hamiltonian to obtain two effective dynamics for the pseudospin and boson variables respectively, where the other variables enter as simple parameters.  This approach will be self consistently valid only if the frequency of the precessing pseudospin $\omega_F=\omega_P(q)$ is much larger than the frequency of the bosonic variables, $\omega_B$, which evolve according to the effective Hamiltonian $H_{m'}(p,q)=\frac{\omega}{2}p^2+V_{m'}(q)$. A detailed analysis of the validity of this approach is presented below.     

An important consequence of this approach is that the dynamics of the semi-classical model and the spectrum in the quantized version, are organized in a finite number of bands, each characterized by the quantum number $m'$. Each band is associated to a different effective Hamiltonian, $H_{m'}$, for the bosonic variables, and since the energy of the effective Hamiltonian is only lower bounded, each band extends infinitely in energy. This band structure provides a useful description of the regular part of the energy spectra, up to a point where the approximation breaks down.  

To determine the values of the fast and slow frequencies ($\omega_{F}$ and $\omega_B$) as a function of coupling and energy, and to be able to establish the range of validity of the previous approach, we analyse the dynamics of the slow bosonic variables and  focus our attention on the effective potential $V_{m'}(p,q)$.   

\begin{figure}
\begin{tabular}{cc}
(a)&(b)\\
\includegraphics[width=0.48\textwidth]{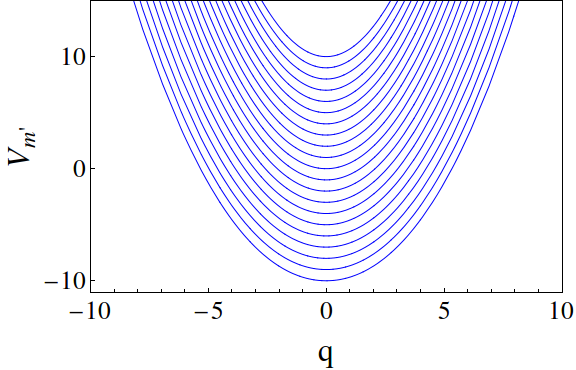}&\includegraphics[width=0.48\textwidth]{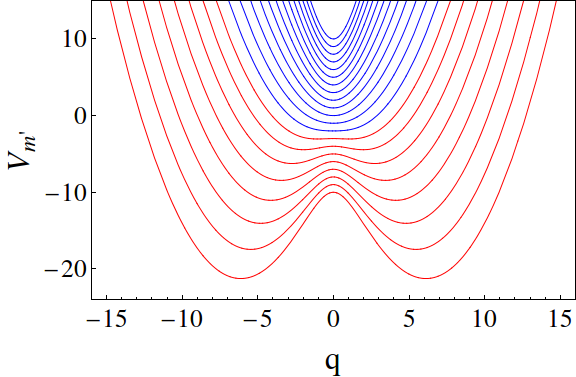}
\end{tabular}
\caption{Effective potential $ V_{m'}(q)$, in the case $\omega/\omega_o=1$ and $j=10$, for the normal (a), $f=\gamma/\gamma_c=0.6$, and superradiant phase (b), $f=\gamma/\gamma_c=2$. From below, the different curves correspond, respectively , to $m'=-10,-9,...,9,10$. Red curves correspond to  potentials with spontaneous breaking of the parity symmetry, which appears only in the superradiant phase}
\label{fig:1}
\end{figure}

The minimum of the potential $V_{m'}(q)$, for $m'$ integer or half integer according to $j$, is given by
\begin{equation}
E_{min}^{m'}=\left\{\begin{array}{cc} -\frac{j\omega_o}{2}\left[ \left(\frac{1}{f}\right)^2+  \left(\frac{m'}{j}\right)^2f^2 \right] & {\mbox{for $-j\leq m'\leq -j/f^2$}}\\
m'\omega_o& {\mbox{otherwise}}
\end{array}\right.
\label{eq:eminm}
\end{equation}
where $f\equiv \gamma/\gamma_c$ with $\gamma_c\equiv\sqrt{\omega\omega_o}/2$, the critical value separating the normal and superradiant phases. The value of $q$ which minimize the potential is
\begin{equation}
q_{min}^{m'}=\left\{\begin{array}{cc} \pm \frac{\omega_o \sqrt{j}}{2\gamma}\sqrt{ \left(\frac{m'}{j}\right)^2f^4-1} & {\mbox{for $-j\leq m'\leq -j/f^2$}}\\
0& {\mbox{otherwise}}
\end{array}\right.
\label{eq:qmin}
\end{equation}
The form of the potentials $V_{m'}(q)$ is illustrated in Fig.\ref{fig:1} for two representative cases in the normal ($\gamma<\gamma_c$) and superradiant ($\gamma>\gamma_c$) phases. In both cases the smaller the value of $m'$, the lower the associated potential. In the normal phase, the potentials for all $m'$ have their minima at $q=0$, whereas in the superradiant phase, the potentials for the values of $m'$ in the interval $-j\leq m'\leq -j/f^2$ present a spontaneous breaking of the parity symmetry ($q\rightarrow -q$), and the potentials  have two degenerate minima. The right-hand side well, with $q>0$, corresponds to a positive rotation angle $\beta$, whereas the left-hand side, with $q<0$, is associated to a negative $\beta$ angle. For $-j/f^2<m'\leq j$ the parity symmetry of the potentials is spontaneously restored and the minima are located again  at $q=0$. 

\subsubsection{Fast Pseudospin approximation: boson and pseudospin frequencies}

For each band $m'$ associated to the potential $V_{m'}(q)$, it is easy to estimate the frequencies of its lowest energy  $E_{min}^{m'}$. For the slow bosonic variables, the frequency is estimated by expanding the potential around the minimum $q_{min}^m$, the quadratic term of this expansion gives the frequency (see \ref{appendixA});  the result is
\begin{equation}
\omega_{B}(E_{min}^{m'})=  \left\{\begin{array}{cc} 
\omega\sqrt{\frac{f^4-\left(\frac{j}{m'}\right)^2}{f^4}}=\omega \sqrt{\frac{2(1+ \epsilon f^2)}{1+2\epsilon f^2}} & {\mbox{for $-j\leq m'\leq -j/f^2$}}\\
\omega\sqrt{1+\frac{m'}{j}f^2}=\omega\sqrt{1+\epsilon f^2}& {\mbox{otherwise}}
\end{array}\right. .
\label{eq:frecBos}
\end{equation}
The expressions in terms of  $\epsilon\equiv E_{min}^{m'}/(\omega_o j),$ were obtained by substituting  $m'$ as a function of $E_{min}^{m'}$ using Eq.(\ref{eq:eminm}). The slow boson frequency for the lowest energy band ($m'=-j$) reduces to
\begin{equation}
\omega_B\equiv\omega_{B}(E_{min}^{-j})=  \left\{\begin{array}{cc} \omega \sqrt{1-f^{-4}} & {\mbox{for $\gamma>\gamma_c$}}\\
\omega\sqrt{1-f^2}& {\mbox{for $\gamma<\gamma_c$}}
\end{array}\right.
\end{equation}

The fast frequencies of the precessing pseudospin are obtained by evaluating $\omega_P$ (\ref{eq:rotation}) at  $q_{min}^{m'}$ (\ref{eq:qmin}), $\omega_F(E_{min}^{m'})=\omega_P(q_{min}^{m'})$, the result is
 \begin{equation}
\omega_F(E_{min}^{m'})=\left\{\begin{array}{cc} \omega_o \left| \frac{m'}{j}\right|f^2=\omega_o\sqrt{-(1+2\epsilon f^2)} 
& {\mbox{for $-j\leq m'\leq -j/f^2$}}  \\
\omega_o& {\mbox{otherwise}}
\end{array}\right.
\label{eq:frecPseu}
\end{equation} 
The fast pseudospin frequency for the lowest energy band  ($m'=-j$) reduces to
\begin{equation}
\omega_F\equiv\omega_{F}(E_{min}^{-j})=  \left\{\begin{array}{cc}  \omega_o f^2 & {\mbox{for $\gamma>\gamma_c$}}\\
\omega_o& {\mbox{for $\gamma<\gamma_c$}}
\end{array}\right.
\end{equation}

\subsubsection{Fast Pseudospin approximation: application range}

\begin{figure}
\begin{tabular}{ccc}
\includegraphics[width=0.3\textwidth]{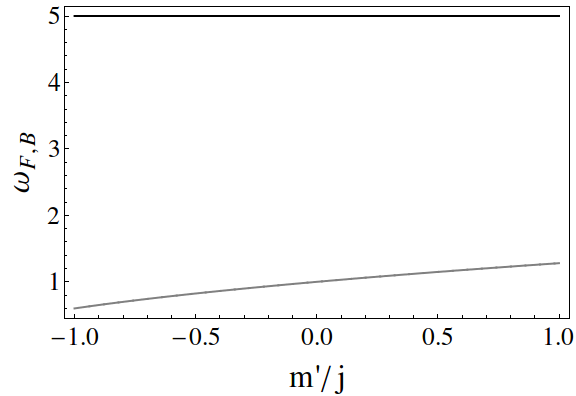}   & \includegraphics[width=0.3\textwidth]{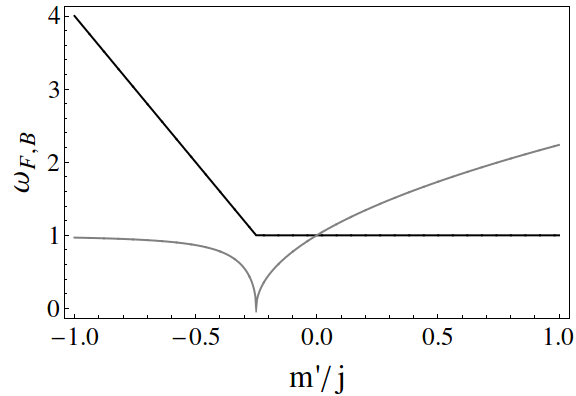}& \includegraphics[width=0.3\textwidth]{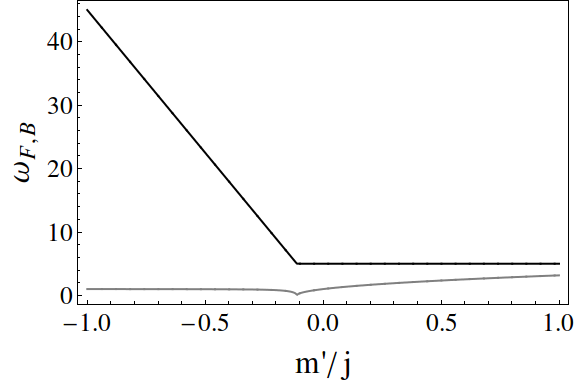}\\  \includegraphics[width=0.3\textwidth]{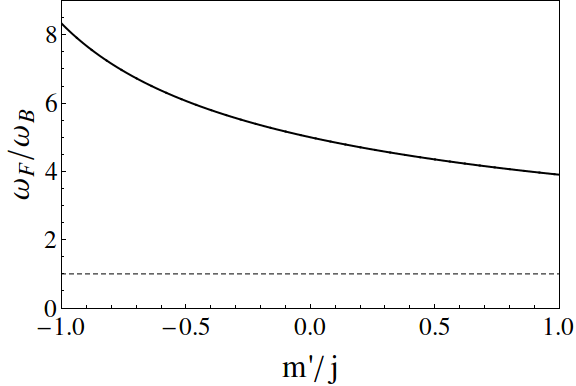}&\includegraphics[width=0.3\textwidth]{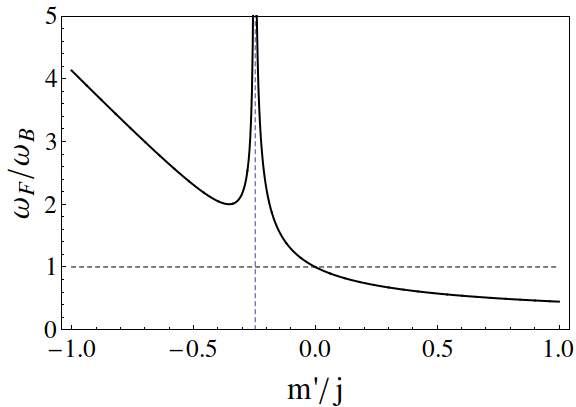}&\includegraphics[width=0.3\textwidth]{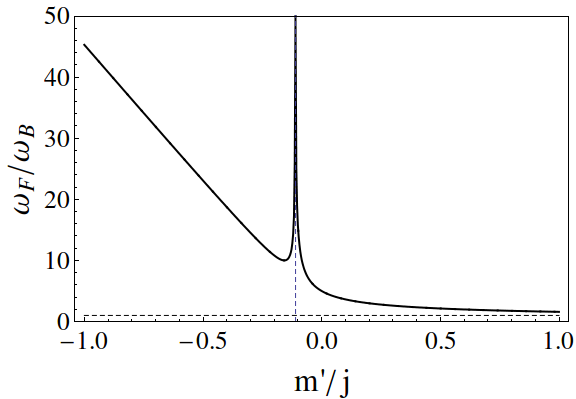}
\end{tabular}
\caption{Top Row: Frequencies $\omega_F(E_{min}^{m'})$ (black lines) and $\omega_B(E_{min}^{m'})$ (gray lines) as a function of $m'/j$. Bottom row  shows the ratio $\omega_F(E_{min}^{m'})/\omega_B(E_{min}^{m'})$.  Left  column: normal phase $\gamma/\gamma_c=0.8$, $\omega/\omega_o=0.2$. Middle column: $\gamma/\gamma_c=2$, $\omega/\omega_o=1$. Right column: $\gamma/\gamma_c=3$, $\omega/\omega_o=0.2$. Horizontal  dashed lines in the bottom panels indicate the value $1$, whereas  the vertical ones in the two rightmost panels indicate the value $-1/f^2$, where the parity symmetry is spontaneously restored.}
\label{fig:2}
\end{figure}

As mentioned above, the fast pseudospin approximation is valid if $\omega_F(E_{min}^{m'})>>\omega_B(E_{min}^{m'})$. 
To illustrate the behaviour of the lowest energy frequencies of each band $m'$ and their respective ratio $\omega_F(E_{min}^{m'})/\omega_B(E_{min}^{m'})$ as a function of $m'$, we have chosen one case in the normal ($f=0.8$, $\omega/\omega_o=1/5$), and  two cases in the superradiant phase ($f=2$, $\omega=\omega_o$ and $f=3$, $ \omega/\omega_o=0.2$). They are shown in Fig.\ref{fig:2}. In the case of the normal phase (left column), the pseudospin frequency is equal to $\omega_o$ for all $m'$, and $\omega_B(E_{min}^{m'})$ increases monotonically with $m'$, consequently the ratio $\omega_F(E_{min}^{m'})/\omega_B(E_{min}^{m'})$ decreases as $m'$  increases. In the superradiant phase,  for $m'$ below $-j/f^2$ the pseudospin frequency descends linearly with $m'$, whereas the boson frequency decreases much more slowly reaching a zero value in $m'\approx -j/f^2$. The maximal ratio $\omega_F(E_{min}^{m'})/\omega_B(E_{min}^{m'})$ is obtained in the minimal value $m'=-j$. For $m'$ above $-j/f^2$ the pseudospin frequency becomes constant and equal to $\omega_o$, but the boson frequency increases with $m'$ making that the ratio $\omega_F(E_{min}^{m'})/\omega_B(E_{min}^{m'})$ decreases as a function of $m'$. 

A divergence in the ratio $\omega_F(E_{min}^{m'})/\omega_B(E_{min}^{m'})$ is observed for  $m'$ around $-j/f^2$ (vertical dashed lines in the figures), which is the value where the parity symmetry is spontaneously restored and the effective potential passes from a double well to a single well. This passage is accompanied by a flatting of the minima of the potential which causes that the boson frequency goes to zero and induces a divergent ratio. Beyond this critical $m'$ value the ratio $\omega_F(E_{min}^{m'})/\omega_B(E_{min}^{m'})$ continues descending as $m'$  increases. Observe that in the $f=3$  out of resonance case (right column), the ratio $\omega_F/\omega_B$ is one order of magnitude greater than in the resonant $f=2$ case, which indicates that the approximation must work much better in the former case, as shown below. 

 As the lower energies are associated with the more negatives values of $m'$, in the three cases the approximation works better in the low energy region, where the  ratio $\omega_F(E_{min}^{m'})/\omega_B(E_{min}^{m'})$ is larger, and ceases to work as the energy increases. The $m'$ value where this happens must be lower for values of $f$ closer to 1 and/or for smaller values of the ratio $\omega_F/\omega_B$. 

\begin{figure}
\centering
\includegraphics[width=0.5\textwidth]{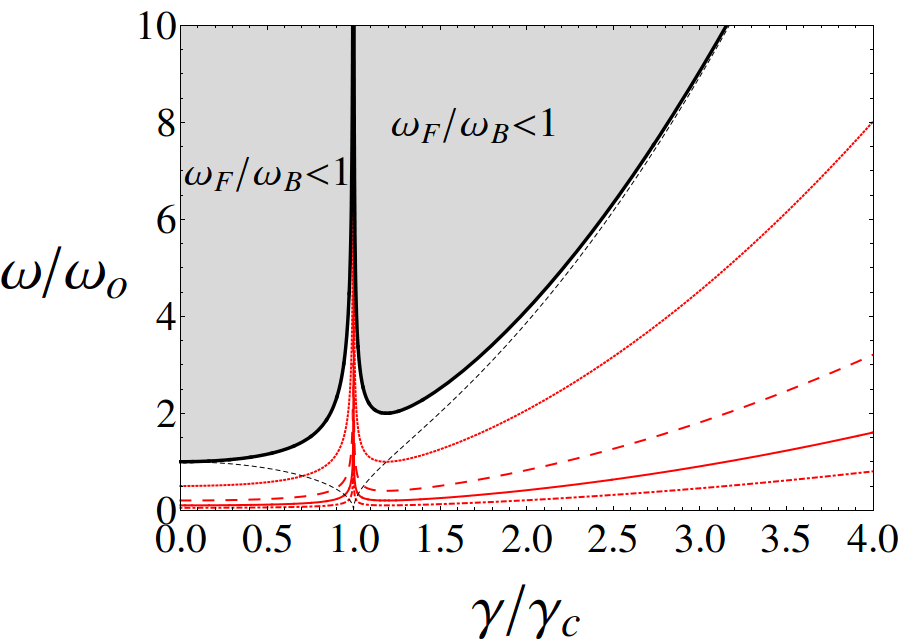}\includegraphics[width=0.5\textwidth]{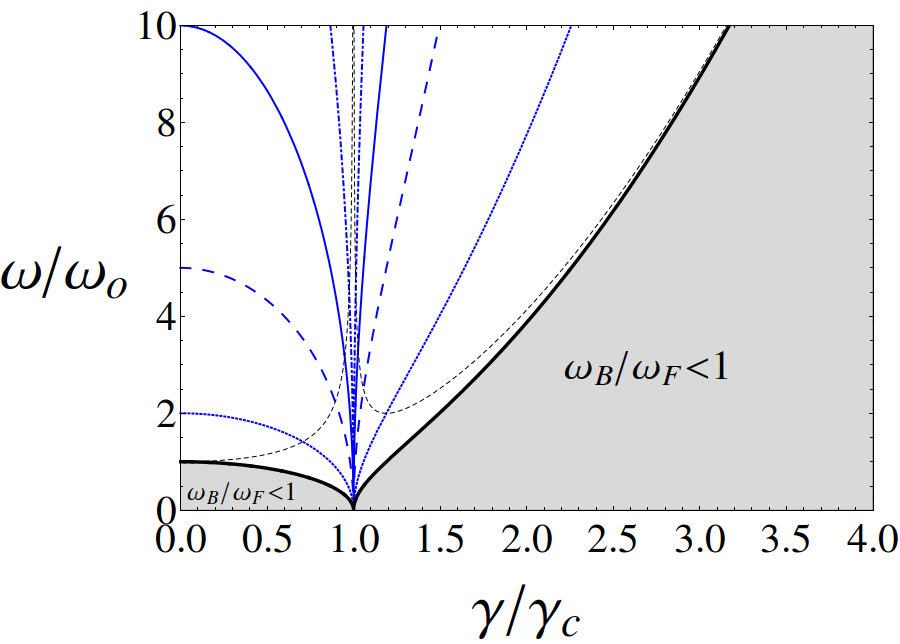}
\caption{Left: Level curves of the ratio $\omega_F/\omega_B$ calculated from the fast pseudospin approximation in the parameter space $\omega/\omega_o$-$\gamma/\gamma_c$. Red dotted, dashed, solid and dot-dashed lines are for  $\omega_F/\omega_B=2,5,10$ and $20$ respectively. The gray zone indicates the  region where the ratio is less than $1$. The thick solid lines is for $\omega_F/\omega_B=1$. Right: similar to left panel but for the ratio   $\omega_F/\omega_B$ calculated from the fast boson approximation. Black dashed lines in every panel indicate the level curve $\omega_F/\omega_B=1$ from the opposite  approximation, fast boson in the left panel and fast pseudospin in the right one.}
\label{fig:4}
\end{figure}

Since the ratio $\omega_F(E_{min}^{m'})/\omega_B(E_{min}^{m'})$ (except for the divergence in the narrow region around $-j/f^2$ in the superradiant phase) decreases as a function of $m'$, a simple and necessary condition for the validity of the fast pseudospin approximation can be established by looking at the frequencies in the lowest  $m'=-j$ band  
\begin{equation}
\frac{ \omega_F(E_{min}^{-j})}{\omega_B(E_{min}^{-j})}=
\left\{\begin{array}{cc}
\frac{\omega_o}{\omega}\frac{f^4}{\sqrt{f^4-1}}&{\mbox{for $\gamma>\gamma_c$}}\\
\frac{\omega_o}{\omega}\frac{1}{\sqrt{1-f^2}}& {\mbox{for $\gamma<\gamma_c$}}
\end{array}\right.,
\end{equation}
the condition $\omega_F/\omega_B>>1$   entails the following condition on the model parameters
\begin{equation}
\frac{\omega}{\omega_o}<<\frac{f^4}{\sqrt{f^4-1}}    {\mbox{\ \ for $\gamma>\gamma_c$\ \ \ and\ \ }}
\frac{\omega}{\omega_o}<< \frac{1}{\sqrt{1-f^2}}{\mbox{\ \ for $\gamma <\gamma_c$}}.
\end{equation}
The level curves of the ratio $\omega_F/\omega_B$ in the parameter space of the Dicke model ($\omega/\omega_o$ vs $\gamma/\gamma_c$) are shown in the left panel of Fig.\ref{fig:4}. The gray area indicates the region where the ratio $\omega_F/\omega_B<1$ and the fast pseudospin is not valid. The solid thick line indicates a ratio $\omega_F/\omega_B=1$. 
Observe that  the  condition $\omega_F/\omega_B<<1$ is fulfilled even  in the resonant case $\omega/\omega_0=1$ for couplings large enough in the superradiant phase, a result  contrary to the naive expectation that the adiabatic approximation with fast pseudospin and slow bosons is only valid if $\omega/\omega_o<<1$. In the normal phase, the condition $\omega_F/\omega_B<<1$  likewise holds for  couplings close to the critical one, for small enough  $\omega/\omega_o$.

However,  the previous condition is a necessary but not sufficient condition to establish the validity of the fast pseudospin approximation. As discussed in section 3, a more restrictive and accurate condition can be established by comparing the frequencies $\omega_{F,B}$ with those obtained from a Holstein-Primakoff or quadratic approximation around the minimal energy configuration of the Dicke model.  

\subsection{Fast boson approximation}

Now we turn our attention to the opposite limit, i.e when the slow variables are the pseudospin ones and the boson variables are fast. To derive in a simple way the approximation, let us consider this time the Dicke Hamiltonian with the pseudospin variables classical and {\it frozen} ($J_i\rightarrow j_i$)
$$
H_B=\omega a^\dagger a+\omega_o j_z+\frac{\sqrt{2}\gamma}{\sqrt{j}} \left(a+a^\dagger\right)j_x,
$$
this Hamiltonian can be easily diagonalized by considering a shift transformation $a=b-\frac{\sqrt{2}\gamma}{\omega\sqrt{j}}j_x$, the resulting Hamiltonian is
$$
H_B=\omega b^\dagger b+\omega_o j_z-\frac{2\gamma^2}{\omega j}j_x^2.
$$
By considering eigenstates of the number operator $ b^\dagger b |n'\rangle=n'|'n\rangle$, we obtain,  for each $n'$, an effective Hamiltonian, $H_{n'}\left(\vec{j}\,\right)$, for the slow classical pseudospin variables
\begin{equation}
H_B|n\rangle=H_{n'}\left(\vec{j}\,\right)|n'\rangle\equiv \left(\omega n'+\omega_o j_z-\frac{2\gamma^2}{\omega j}j_x^2\right)|n'\rangle.
\label{Hnp}
\end{equation}
The resulting Hamiltonian for the pseudospin variables, $H_{n'}\left(\vec{j}\,\right)$ is a Lipkin-Meshkov-Glick one (as already noticed in Refs.\cite{Keeling10,Itin10}), whose  energy minimum  is
\begin{equation}
E_{min}^{n'}=\left\{\begin{array}{cc}-\omega_o j+\omega n' & {\mbox{for $\gamma<\gamma_c$}}\\
-\frac{\omega_o j}{2}\left(f^2+\left(\frac{1}{f}\right)^2\right)+\omega n' & {\mbox{for $\gamma\geq \gamma_c$}}
\end{array}\right.
\end{equation}
with the variables minimizing the energy given by 
\begin{equation}
(j_{x,min},j_{y,min},j_{z,min})=\left\{\begin{array}{cc}(0,0,-j) & {\mbox{for $\gamma<\gamma_c$}}\\
(\pm j\sqrt{1-f^{-4}},0,-\frac{j}{f^2},) & {\mbox{for $\gamma\geq \gamma_c$}}
\end{array}\right.
\label{eq:minconFB}
\end{equation}

\begin{figure}
\begin{tabular}{cc}
(a)&(b)\\
\includegraphics[width=0.48\textwidth]{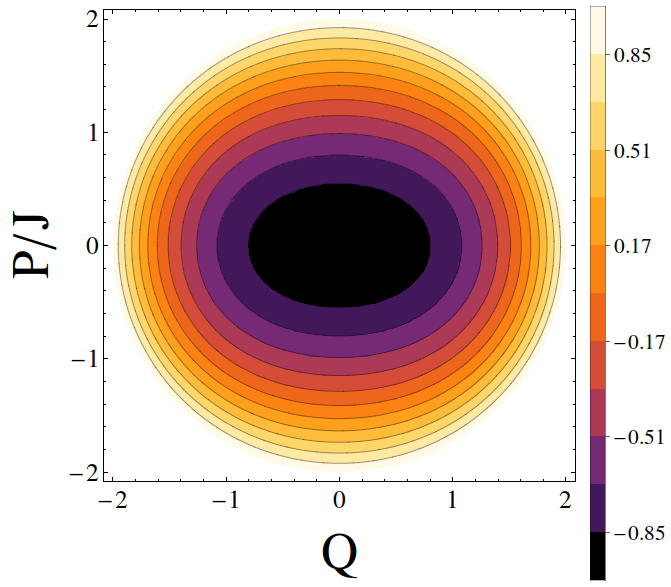}&\includegraphics[width=0.48\textwidth]{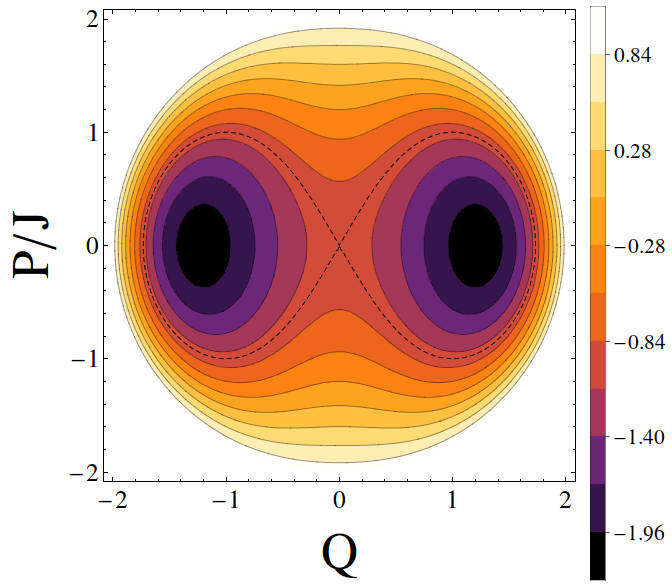}
\end{tabular}
\caption{Equipotential lines of the effective fast-boson Hamiltonian $(H_{n'}-\omega n')/j$, in terms of the canonical variables $Q$ and $P$ defined in the main text. The panel (a) is for $f=0.8$ and (b) for $f=2$. In both cases a ratio $\omega/\omega_o=10$ was used.}
\label{fig:3}
\end{figure}

The phase space of the effective Hamiltonian for the pseudospin slow variables, $H_{n'}\left(\vec{j}\,\right)$, is  the surface of a sphere of radius $j$. To illustrate its behavior, we show in Fig. \ref{fig:3}  contour plots of the equipotential lines of
 $H_{n'}\left(\vec{j}\right)$ (without the additive term $\omega n'$), using the canonical variables 
\begin{eqnarray}
Q=\sqrt{2(1+j_z/j)}\cos\phi \nonumber\\
P=-j \sqrt{2(1+j_z/j)}\sin\phi,
\label{eq:cano}
\end{eqnarray} 
where $\phi$ is the azimuthal angle defined through $\tan\phi=j_y/j_x$. Two illustrative cases are presented,  for  couplings below and above the critical one and for a frequencies ratio $\omega/\omega_o=10$.  

The spontaneous breaking of the parity symmetry is seen by the appearance of a double degenerated minima for couplings above the critical value. It is worth mentioning that, contrary to the fast pseudospin approximation, in this case the spontaneous breakings of the parity symmetry and the accompanied quantum phase transitions occurs simultaneously for all the effective potential $H_{n'}\left(\vec{j}\,\right)$ when the coupling reaches the critical value $\gamma_c$. The term in the effective potential, depending on the fast boson quantum number $\omega n'$, is a simple additive constant without effect in the quantum phase transition, which is in contrast with the fast pseudospin approximation where the quantum number of the fast variable ($m'$) determines, for the superradiant phase,  if the associated effective potential for the slow variable presents or not a spontaneous breaking of the parity symmetry  [see panel (b) of Fig.\ref{fig:1}].       

Note that the fast boson approximation implies  the existence of an infinite number of bands in  the corresponding quantum energy spectrum. Each band labelled by the quantum number $n'$ is associated to an effective Hamiltonian for the pseudospin variables $H_{n'}\left(\vec{j}\right)$. Since this effective Hamiltonian is lower and upper bounded in energy, every band has a finite extension in energy. Additionally, observe that the pseudospin dynamics is independent of the quantum number $n'$, whose  only effect  is an energy  shifting   by an amount $\omega n'$.

\subsubsection{Fast boson approximation: boson and pseudospin frequencies}

In order to determine the region of applicability of the previous fast boson approximation, the frequencies of the boson  and pseudospin variables have to be evaluated. A necessary  condition of applicability of the fast boson approximation is $\omega_B>>\omega_F$.

For the bosonic variables the frequency is simply the frequency of the non-interacting case
$$
\omega_B=\omega, 
$$
regardless of the value of the coupling. 
For the slow  pseudospin variable, we  calculate  the frequency for the minimal energy of the effective Hamiltonian $H_{n'}$ by expanding it up to second order around the minimal energy configuration Eq.(\ref{eq:minconFB}). The result (see \ref{appendixA} for details) is independent of the quantum number $n'$ and is given by
\begin{equation}
\omega_{F}(E_{min}^{n'})=  \left\{\begin{array}{cc}  \omega_o\sqrt{f^4-1} & {\mbox{for $\gamma>\gamma_c$}}\\
\omega_o\sqrt{1-f^2}& {\mbox{for $\gamma<\gamma_c$}}
\end{array}\right.
\end{equation} 

\subsubsection{Fast boson approximation: application range}

With the previous expression for the frequencies, the ratio $\omega_B/\omega_F$ is
\begin{equation}
\frac{\omega_B}{\omega_F}=\left\{\begin{array}{cc}
\frac{\omega}{\omega_o}\frac{1}{\sqrt{f^4-1}}&{\mbox{for $\gamma>\gamma_c$}}\\
\frac{\omega}{\omega_o}\frac{1}{\sqrt{1-f^2}}& {\mbox{for $\gamma<\gamma_c$}}
\end{array}\right.,
\end{equation}
therefore  the condition $\omega_B/\omega_F>>1$ entails
\begin{equation}
\frac{\omega}{\omega_o} >> \sqrt{1-f^2} {\mbox{\ \ \ for $\gamma<\gamma_c$\ \ \  and \ \ \ }}
 \frac{\omega}{\omega_o} >> \sqrt{f^4-1}{\mbox{\ \ \ for $\gamma>\gamma_c$}}.
\end{equation}
The level curves of the ratio $\omega_B/\omega_F$ in the relevant parameter space of the Dicke model are shown in panel (b) of Fig.\ref{fig:4}. Similar to the fast pseudospin approximation, the region where the fast boson approximation is not valid ($\omega_B/\omega_F<1$) is indicated by the gray area.

The  previous necessary but rough condition for the validity of the fast boson approximation can be combined with  the previously obtained for the fast pseudospin one. Observe, however, that there exists a region in the parameter space, the region between the dashed black and thick black lines in both panels of Fig.\ref{fig:4}, where none of the fast-slow approximations is ruled out by the simple conditions $\omega_F/\omega_B>1$ or $\omega_B/\omega_F>1$. This would imply that in this region both approximations would be, in a certain extent, valid, which is clearly impossible. In order to overcome this contradiction a more restrictive condition for the validity of one or another fast-slow approximation has to be established. This more restrictive condition can be obtained  by comparing the above derived frequencies, $\omega_{B,F}$ with those obtained from a Holstein-Primakoff approximation \cite{Bran03} as  is discussed in the following section.

\section{Tighter applicability limits of the fast-slow approximations}

\begin{figure}
\begin{tabular}{cc}
(a)&(b)\\
\includegraphics[width=0.48\textwidth]{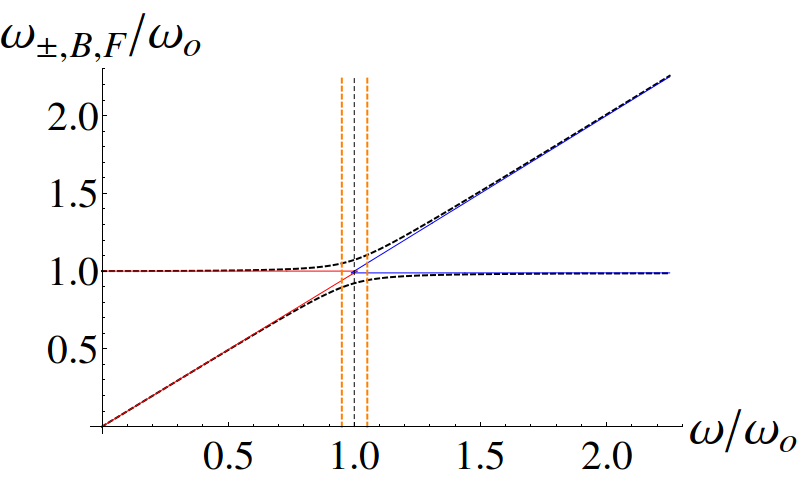}&\includegraphics[width=0.48\textwidth]{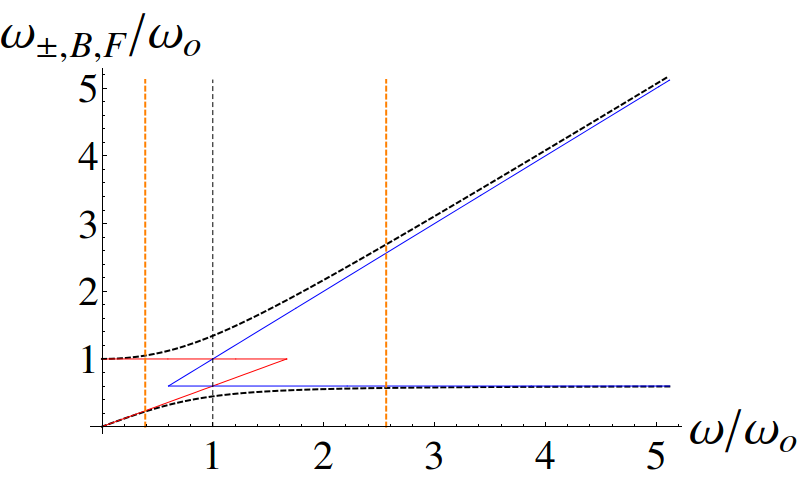}\\
(c)&(d) \\
\includegraphics[width=0.48\textwidth]{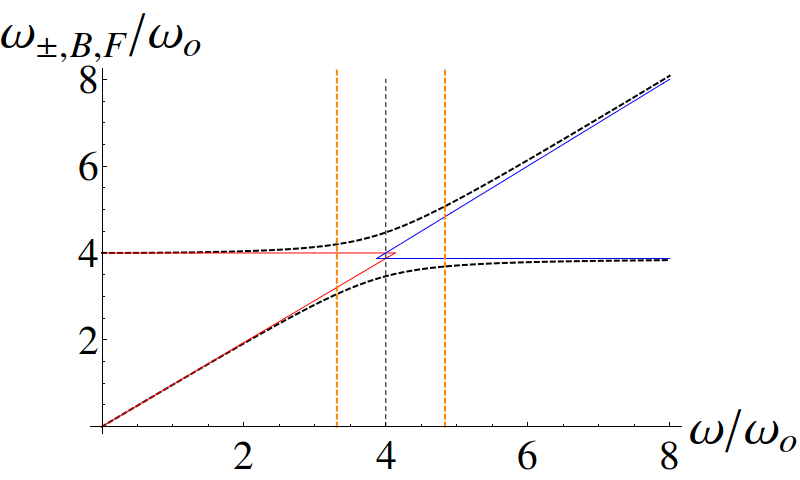}&
\includegraphics[width=0.48\textwidth]{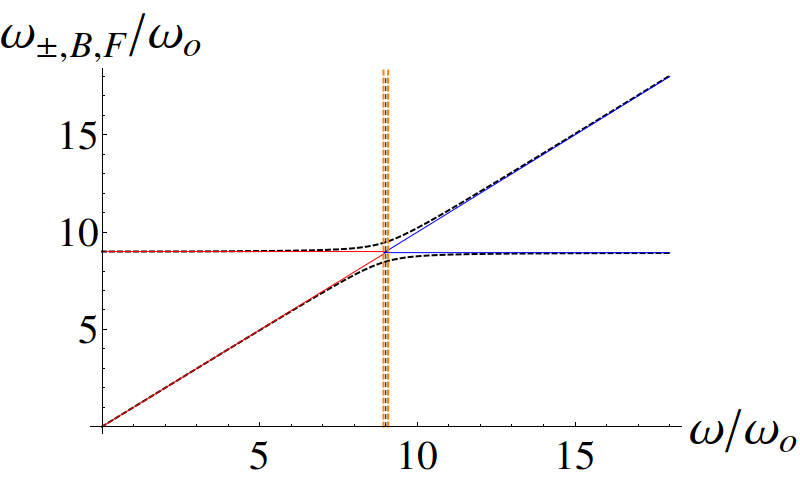}
\end{tabular}
\caption{Quadratic (or Holstein-Primakoff) frequencies $\omega_\pm$ (dashed lines), compared with the  fast pseudospin  (red lines) and fast boson (blue lines)  approximation frequencies ($\omega_{F,B}$), as a function of  $\omega/\omega_o$ for fixed values of the ratio $f=\gamma/\gamma_c$: (a) $f=0.15$, (b) $f=0.8$, (c)$f=2$ and  (d) $f=3$. In the interval $\omega/\omega_o$ between the vertical orange lines, the fast-slow approximation frequencies differ for more than $5\%$  from the quadratic ones. The vertical black dashed line is the $\omega/\omega_o$ value where the fast pseudospin frequencies intersect the fast boson ones, this value is $1$ for the normal phase and $f^2$ in the superradiant one.}
\label{fig:5}
\end{figure}

\begin{figure}
\centering
\includegraphics[width=0.8 \textwidth]{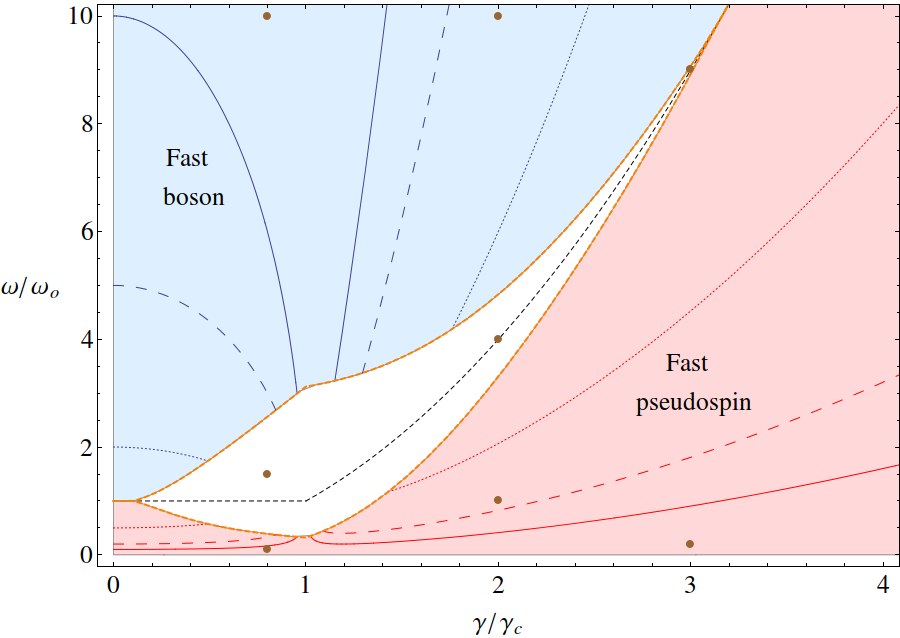}
\caption{Region of validity of the fast-slow approximations in the $\omega/\omega_o$ vs $\gamma/\gamma_c$ space. Red  and blue regions indicate, respectively,  the region of validity of the fast pseudospin and fast boson approximations. The white, orange bounded, area indicates the region where no approximation is expected to be valid because  the frequencies of both approximations differ (for more than 5 \%) from the quadratic ones $\omega_\pm$. Dotted, dashed and solid lines are level curves ($2$, $5$ and $10$ respectively) for $\omega_F/\omega_B$ in the red area and for $\omega_B/\omega_F$ in the blue one. The larger these frequencies ratios, the better the respective approximation is expected to work. The dashed black lines in the white region corresponds to $1$ for $f=\gamma/\gamma_c<1$ and $f^2$ for $f\geq 1$.  The  brown points coordinates,  indicating the cases shown in Figs.\ref{fig:7}, \ref{fig:8} and \ref{fig:9}, are: $(\gamma/\gamma_c,\omega/\omega_o)=(0.8,0.1),(0.8, 1.5), (0.8, 10), (2, 1), (2, 4), (2, 10), (3, 0.2), (3, 9)$.}
\label{fig:6}
\end{figure} 

While in the previous section a simple condition on the parameters of the model was established to determine the region of validity of one or another fast-slow approximation, a tighter and consistent 
condition can be derived by comparing the frequencies coming from the fast-slow approximations with those ($\omega_{\pm}$) obtained from a Holstein-Primakfoff approximation to the Dicke model\cite{Bran03} , which are the same that are obtained by considering a quadratic approximation around the minimal energy configuration in the exact Dicke model. These frequencies are given by
$$
2\omega_{\pm}^2=\omega^2+\omega_o^2\pm\sqrt{(\omega_o^2-\omega^2)^2+16 \gamma^2\omega\omega_o}
$$
for $\gamma<\gamma_c$ and by 
$$
2 \gamma_c^4 \omega_{\pm}^2=\omega_o^2\gamma^4+\omega^2\gamma_c^4\pm\sqrt{(\omega_o^2\gamma^4-\omega^2\gamma_c^4)^2+4\omega^2\omega_o^2\gamma_c^8}.
$$
for the superradiant phase $\gamma>\gamma_c$. 

These frequencies $\omega_\pm$ provide the  exact  low energy frequencies  in the thermodynamic limit, except for the critical coupling ($\gamma=\gamma_c$) where one of the quadratic terms is exactly zero and the lowest order approximation involves a quartic term. We naturally demand that the frequencies obtained from our fast-slow approximations reproduce this low energy limit. 

In Fig.\ref{fig:5} we compare the frequencies $\omega_\pm$ (dashed) with $\omega_F$ and $\omega_B$ coming from the fast pseudospin (red) and fast boson (blue) approximation. The curves show the frequencies (in units of $\omega_o$) as a function of the ratio $\omega/\omega_o$ for fixed values of the ratio $f=\gamma/\gamma_c$ in the normal ($f=0.15$ and $f=0.8$) and superradiant  ($f=2$ and $f=3$) phases. Except for a region around  $\omega/\omega_o=1$ in  the normal phase and around $\omega/\omega_o=f^2$ in the superradiant one, the frequencies coming from the fast-slow approximations reproduce very well the frequencies $\omega_\pm$. The regions between the vertical dashed orange lines correspond to the ratios $\omega/\omega_o$ where the $\omega_{B,F}$ frequencies differ from the $\omega_\pm$ ones for more than 5 \%.  Therefore, these regions are excluded from the region of validity of any fast-slow approximation. Observe that these excluded regions  are equally those where the ratio $\omega_B/\omega_F\sim 1$. Summarizing, we have identified  regions in the parameter space of the Dicke model, where the conditions  $\omega_B/\omega_F>>1$ (fast boson approximation) or $\omega_F/\omega_B>>1$ (fast pseudospin approximation) are fulfilled and, additionally,   the frequencies $\omega_{B,F}$ reproduce very well the Holstein-Primakoff ones $\omega_{\pm}$. 

In  Fig.\ref{fig:6}, the validity analysis described above is extended to all the couplings in the interval $\gamma_/\gamma_c\in[0,4]$. The region of validity of the fast pseudospin approximation is indicated in light red, whereas the region where the fast boson is valid is indicated in light blue. The white region indicates the region in the parameter space where none of the two approximation is valid, in this region the frequencies $\omega_{F,B}$ differ from the normal frequencies $\omega_\pm$ in more than 5 \% as discussed above. The Fig.\ref{fig:6} is a conclusive and very concrete result that gives a valuable guide to determine if  given values of the Dicke Hamiltonian parameters produce a Hamiltonian where one,  another or neither fast-slow approximation can be used. In the same figure different level curves for the ratio of boson and pseudospin frequencies are drawn. In the upper light blue region the curves correspond to $\omega_B/\omega_F=2$ (dotted), $\omega_B/\omega_F=5$ (dashed) and $\omega_B/\omega_F=10$ (solid) respectively. In the bottom light red region similar curves are shown but for the ratio $\omega_F/\omega_B$. The larger these ratios, the more accurate the respective approximation is expected to be (fast boson approximation in the upper blue region and the fast pseudospin one for the lower red region).          

Some important observations, 
\begin{enumerate}
\item In the normal phase, the resonant case $\omega/\omega_o=1$ is  excluded from the region of validity of any fast-slow approximation, but in the superradiant phase  it can be described with
the fast pseudospin approximation and  the greater the coupling, the better  the approximation is expected to work.
\item Although we have imposed the condition  $\omega_{B,F}\sim\omega_\pm$ for the lowest energy states, this does not mean that both approximations are equivalent. As we show numerically below, the fast-slow approximations give a better approximation to the exact Dicke solution and extends until energies well beyond the energies where the Holstein-Primakoff approximation breaks down. 
\item The present analysis for the validity of any fast-slow approximation is based on the frequencies coming from the respective approximations calculated in the lowest (ground state) energy. For  given values of the Hamiltonian parameters,  the validity of the approximations as a function of the energy for a given band ($m'$ in the fast boson or $n'$ in the fast boson approximation) are determined numerically in the next section. They indicate that the larger the ratio $\omega_F/\omega_B$ ($\omega_B/\omega_F$) the more extended in excited energy the fast pseudospin (fast boson) approximation is. 
\item A very simple criterion to establish the validity of one or other approximation can be obtained; in the normal phase ($f<1$) the criterion is $\omega/\omega_o>>1$ for the fast boson BOA whereas for the fast pseudospin BOA the criterion is $\omega/\omega_o<<1$. In the superradiant phase the criterion changes as a consequence of the different nature of the fundamental effective excitations, the simple criterion is now $\omega/\omega_o>> f^2$ and $\omega/\omega_o<< f^2$ for the fast boson and fast pseudospin BOA's respectively. The line $1$ for $f<1$ and $f^2$ for $f>1$ marking roughly the border between one and other BOA's  is indicated in Fig.\ref{fig:6} by dashed black lines.           
\end{enumerate} 

\section{Semiclassical expectation values}

In the previous sections we introduced the two approximated, semiclassical Hamiltonians which allow to decouple the fast and slow modes and determined the  characteristic frequencies associated with these modes and their range of validity. In this section we present the calculation of the expectation values of selected observables ($\mathcal{O}$) in stationary states. 
  
For the fast pseudospin approximation the expectation value  of any observable $\mathcal{O}\left(a,a^\dagger,\vec J\, \right)$ in a stationary state of energy $E$ can be semiclassically evaluated by
\begin{equation}
  \left< {\mathcal O} \right>=\frac{\int \, dp dq \, {\mathcal O}_{m'} (p,q) \delta \left[ E - H_{m'}(p,q) \right]}{\int \, dp dq \,\delta \left[ E - H_{m'}(p,q) \right]},
\label{eq:semi}
\end{equation}
where ${\mathcal O}_{m'} (p,q) = \left< j, m' \right| {\mathcal
  O}(a^{\dagger},a; J_x, J_y, J_z) \left| j,m' \right>$ with $|j,m'\rangle$  eigenvectors of the rotated operator $J_{z'}$,  $H_{m'}(p,q)$ is the effective Hamiltonian for the slow boson variables, and  the bosonic operators $\left( a^{\dagger}, a \right)$ have to be  written in terms of their classical limit variables $(q,p)$:  $a\rightarrow\frac{1}{\sqrt{2}}(q+ip)$ and $a^\dagger\rightarrow\frac{1}{\sqrt{2}}(q-ip)$.

For the other case, the fast boson approximation, the semiclassical approach to the expectation values in stationary states is
\begin{equation}
  \left< {\mathcal O} \right>=\frac{\int \, dj_z d\phi \, {\mathcal O}_{n'} (j_z,\phi) \delta \left[ E - H_{n'}\left(\vec{j}\,\right) \right]}{\int \, dj_z d\phi \,\delta \left[ E - H_{n'}\left(\vec{j}\,\right) \right]},
\label{eq:semi2}
\end{equation}
where ${\mathcal O}_{n'} (j_z,\phi) = \left< n' \right| {\mathcal
  O}(a^{\dagger},a; J_x, J_y, J_z) \left| n' \right>$ with  $|n'\rangle$ eigenvectors of the shifted boson number  operator $b^\dagger b$,  $H_{n'}\left(\vec{j}\,\right)$ is the effective Hamiltonian for the slow pseudospin variables, and  the pseudospin operators are written in terms of  classical canonical variables ($j_z,\phi$): $(J_x,J_y,J_z)\rightarrow (j\sqrt{1-(j_z/j)^2}\cos\phi, j\sqrt{1-(j_z/j)^2}\sin\phi,  j_z)$.
  
We particularize the previous general expressions for two observables, the number of photons $a^\dagger a$ and the population difference between the two level systems $J_z$. The details of the calculation yielding to the given expressions can be found in \ref{appendixB}. 

\subsection{Density of States and semiclassical expectation values: fast pseudospin}

First, we calculate the expression appearing in the denominator of Eq.(\ref{eq:semi}), which  is proportional to the  semiclassical approximation to the Density of States [$\nu^f(E,m')$] or Weyl's formula
\begin{equation}
\fl
2\pi \nu^{f}(E,m')=\int\int dp\,dq\,\,\delta\left[E-H_{m'}(q,p)\right]=\sqrt{\frac{2}{\omega}}\int_{q\in \mathcal{I}_{E,m'}}\,\frac{dq}{\sqrt{E-V_{m'}(q)}},
\label{eq:denFP}
\end{equation}
where $\mathcal{I}_{E,m'}$ is the classical allowed region in $q$  [$E-V_{m'}(q)\geq 0$]. For energies inside the double well potential this region is formed by two disconnected intervals, whereas for any other case is given by a sole interval (see  Fig.\ref{fig:1}) 
$$
\mathcal{I}_{E,m'}=\left\{\begin{array}{cc}
 [-q_-(E,m'),-q_+(E,m')] \cup [q_+(E,m'),q_-(E,m')]& {\mbox{for $-j\leq m'\leq -j/f^2$}}\\
\ &  \mbox{and $E_{min}^{m'}<E\leq \omega_o m'$} \\ 
\  [- q_-(E,m'),q_-(E,m') ]& {\mbox{otherwise}}  
\end{array}\right.
$$
where  $q_\pm$ are the returning points  ($V_{m'}(q)=E$) given by
$$
q_\pm(E,m')=\sqrt{\frac{2j\omega_o}{\omega}}\sqrt{\frac{E}{\omega_o j}+\left(\frac{m'}{j}\right)^2 f^2\pm \frac{m'}{j}\sqrt{1+2\frac{E}{\omega_o j} f^2+\left(\frac{m'}{j}\right)^2 f^4}}.
$$
 For the expectation value of the number of bosons, the results is 
\begin{eqnarray}
 \langle a^\dagger a\rangle_{E,m'}&=\frac{1}{2\pi \nu^{f}(E,m')}
\int dp dq \frac{1}{2}\left(p^2+q^2\right)\delta[E-H_{m'}(p,q)]\nonumber\\
 &= \frac{1}{2\pi \nu^{f}(E,m')}\frac{1}{\omega}\sqrt{\frac{2}{\omega}}\int_{\mathcal{I}_{E,m'}} \frac{E-\sqrt{w_o^2+(\frac{2\gamma q}{\sqrt{j}})^2}}{\sqrt{E-V_{m'}(q)}}dq,
\label{eq:aaFP}
\end{eqnarray}
whereas for the $J_{z}$ operator we obtain 
\begin{eqnarray}
\langle J_z \rangle_{E,m'}&=\frac{1}{2\pi\nu^{f}(E,m')}
\int dp dq \langle j m'| J_z |j,m'\rangle\delta[E-H_{m'}(p,q)]
\nonumber\\
 &=\frac{1}{2\pi\nu^{f}(E,m')}\sqrt{\frac{2}{\omega}}\int_{\mathcal{I}_{E,m'}}\frac{\omega_o\ m' dq}{\sqrt{w_o^2+(\frac{2\gamma q}{\sqrt{j}})^2}\sqrt{E-V_{m'}(q)}}. 
\label{eq:jzFP}
\end{eqnarray}

\subsection{Density of States and semiclassical expectation values: fast boson}

Similarly,  we evaluate first  the denominator of the general formula (\ref{eq:semi2}), which is proportional to the semiclassical approximation to the  energy density of states [$\nu^{a}(E,n')$] 
\begin{equation}
\fl
2\pi \nu^{a}(E,n')=\int d j_z d\phi \delta\left[E-H_{n'}\left(\vec{j}\right)\right]
=\left\{\begin{array}{cc}\frac{8}{\omega_{0}}\displaystyle\int_0^{\phi_o}\frac{\,d\phi}{\sqrt{\mathcal{F}(\cos\phi)}},&{\mbox{ for $\epsilon_{n'}\leq-1$}}\\
\frac{1}{\omega_{0}}\displaystyle\int_0^{2\pi}\frac{\,d\phi}{\sqrt{\mathcal{F}(\cos\phi)}},&{\mbox{ for $\epsilon_{n'} >-1$}}
\end{array}\right.
\label{eq:denFB}
\end{equation}
where we have   defined  $\mathcal{F}(x)=1+2f^{2}\epsilon_{n'} x^2+f^{4}x^{4}$ with  $\epsilon_{n'}\equiv (E-\omega n')/(\omega_o j)$, and  $\phi_o$ is given by $ \cos^{2}(\phi_o)=\frac{1}{f^2}\left[-\epsilon_{n'}+\sqrt{\epsilon_{n'}^{2}-1}\,\right]$.

The expectation value of $J_{z}$, 
$$
\langle J_z\rangle_{n',E}=\frac{1}{2\pi\nu^a(E,n')}\int dj_zd\phi j_z\delta\left[E-H_{n'}\left(\vec{j}\right)\right],
$$
 is given by 
\begin{equation}
\langle J_z\rangle_{n',E}=\left\{\begin{array}{cc}-\frac{8j}{f^2\omega_{0} 2\pi\nu^{a}(E,n')}\displaystyle\int_0^{\phi_o}\frac{\phi\,d\phi}{\cos^2\phi\sqrt{\mathcal{F}(\cos\phi)}},&{\mbox{ for $\epsilon_{n'}\leq-1$}}\\
 & \\
-\frac{j}{f^{2}\omega_{0}2\pi\nu^{a}(\epsilon,n')}\displaystyle\int_0^{2\pi}\frac{\,d\phi \left(1-\sqrt{\mathcal{F}(\cos\phi)}\right)}{\cos^2(\phi)\sqrt{\mathcal{F}(\cos\phi)}},&{\mbox{ for $\epsilon_{n'} >-1$}}
\end{array}\right.
\label{eq:jzFB}
\end{equation}
Finally, for the expectation value of the number of bosons,
$$
\langle a^\dagger a\rangle_{E,n'}=
\frac{1}{2\pi\nu^{a}(E,n')}
\int d j_z d\phi \langle n'|a^\dagger a|n'\rangle  \delta \left[E-H_{n'}\left(\vec{j}\right)\right],
$$
the result is 
\begin{equation}
\fl
\langle a^\dagger a\rangle_{E,n'}=\left\{\begin{array}{cc}
n-\frac{8j}{f^2 2\pi\nu^{a}(E,n')}\displaystyle\int_0^{\phi_o}d\phi \frac{1+\cos^2\phi \epsilon_{n'} f^2}{\cos^2\phi \sqrt{\mathcal{F}(\cos\phi)}} & {\mbox { \ \ \ for $\epsilon_{n'}\leq-1$}}\\
n-\frac{j}{f^2 2\pi\nu^{a}(E,n')}\displaystyle\int_0^{2\pi}d\phi \frac{1+\epsilon_{n'} f^2\cos^2\phi-\sqrt{\mathcal{F}(\cos\phi)}}{\cos^2\phi\sqrt{\mathcal{F}(\cos\phi)}}
&{\mbox{\ \ \ for $\epsilon_{n'}>1$}}
 \end{array}\right.
\label{eq:aaFB}
\end{equation}

\section{Comparison with the exact numerical results}

In this section we compare the results for the observables obtained employing the fast-slow approximations with those evaluated with numerical diagonalization of the exact Dicke Hamiltonian. 

\subsection{Peres lattices and fast-slow approximations}


The exact numerical full quantum Dicke results are presented using Peres lattices \cite{Peres84,Stransky09}, which are built plotting the expectation value of a given observable in the Hamiltonian eigenstates against the corresponding energy eigenvalue. To make the comparisons between the exact and approximate expectation values, we select nine different sets of parameters. Three different ratios $f=\gamma/\gamma_c$ are chosen, one in the normal  ($f=0.8$) and two in the superradiant phase ($f=2$ and  $f=3$). For each value $f$, three different ratios $\omega/\omega_o$ are used. One where the fast pseudospin approximation is expected to work, other where the valid approximation is the fast boson one, and a third ratio where neither the fast pseudospin nor the fast boson approximation are valid. Concretely, for $f=0.8$ the ratios $\omega/\omega_o$ used are $0.1$, $1.5$ and $10$. For $f=2$ we use the ratios $\omega/\omega_o=1,4$ and $10$, and finally, for $f=3$ we use $\omega/\omega_o=0.2,9$ and $20$. All these points (excepting the last one that is out of the plot range) are indicated in Fig.\ref{fig:6} by brown dots. We choose Peres latices for observable $J_z$ to make the comparisons.

\begin{figure}
\begin{tabular}{cc}
(a)&(b)\\
\includegraphics[width=0.46\textwidth]{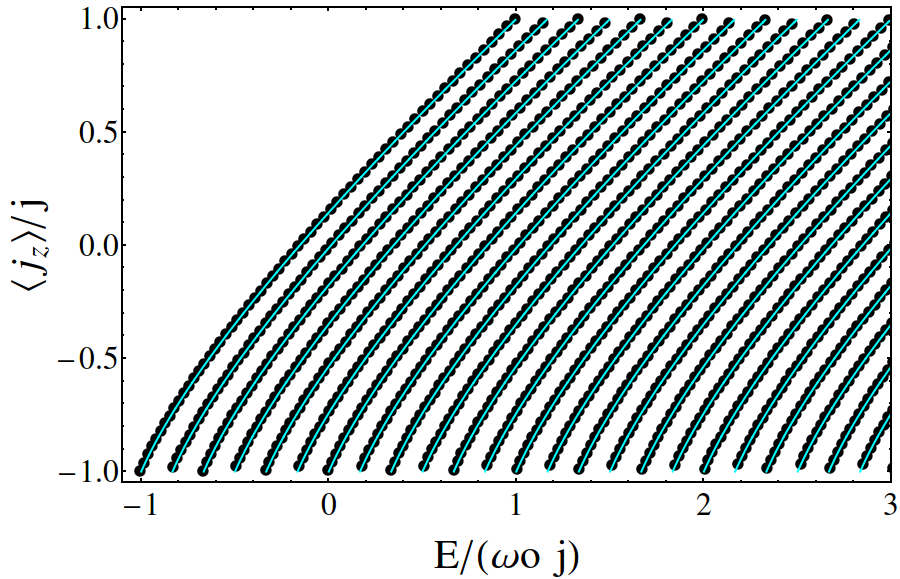}&\includegraphics[width=0.48\textwidth]{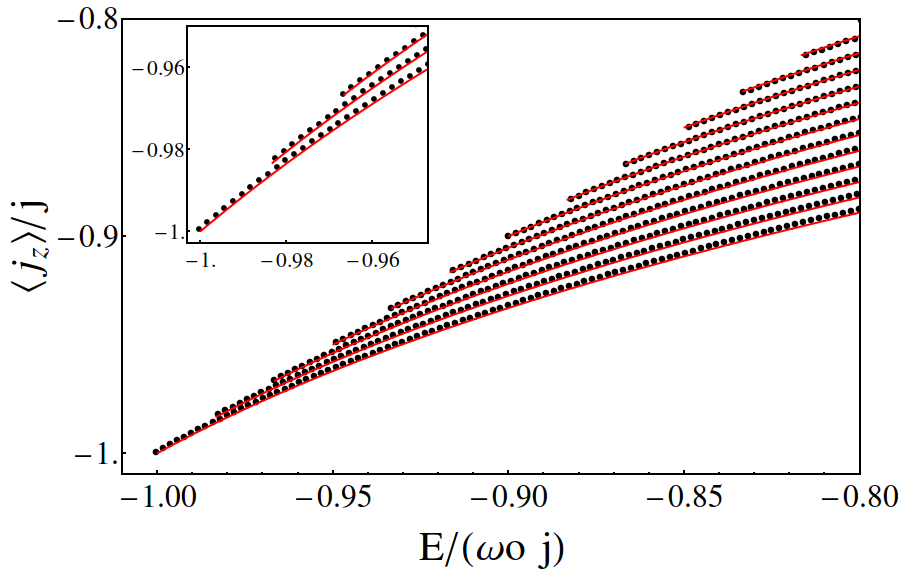}\\ 
(c)&(d)\\
\includegraphics[width=0.47\textwidth]{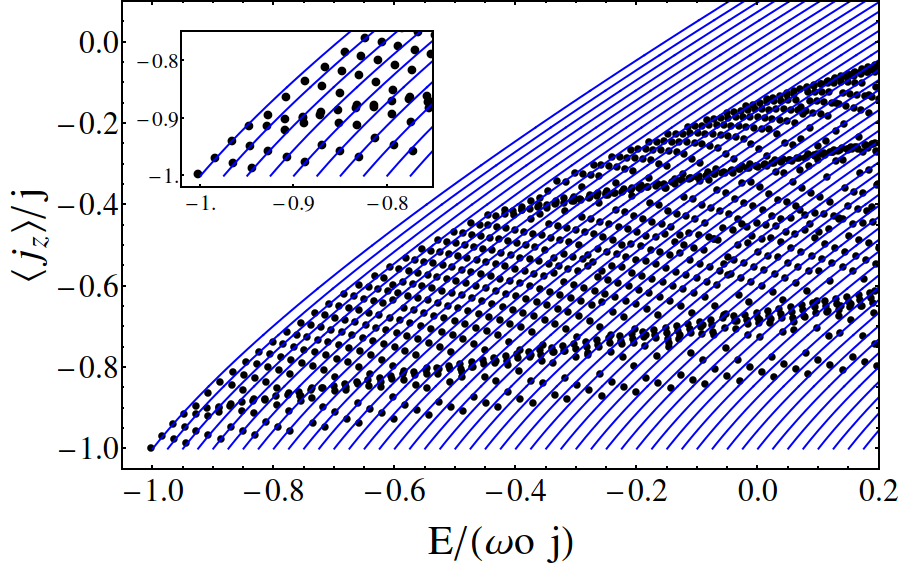}&\includegraphics[width=0.47\textwidth]{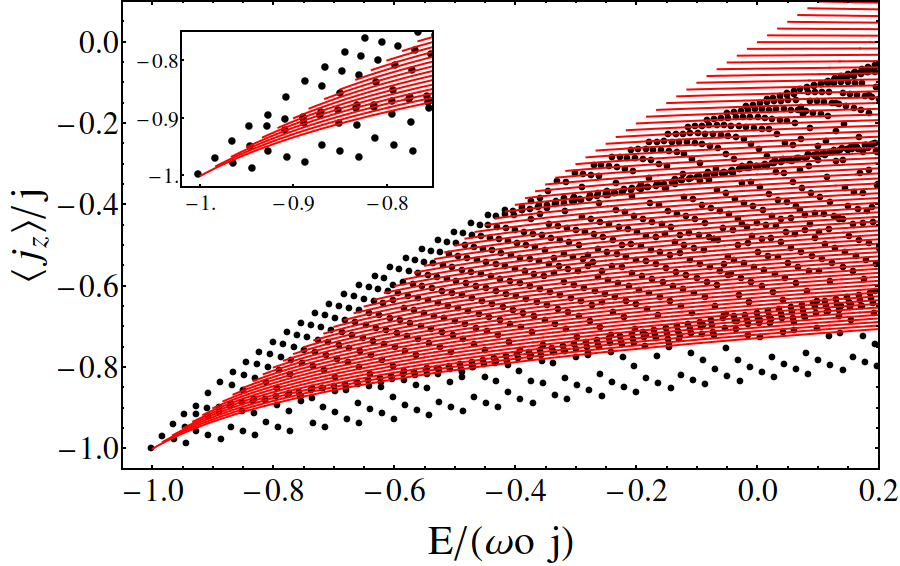}\\
\end{tabular}
\caption{Peres lattices $\langle J_z \rangle/j$ vs $E/(\omega_o j)$ (black dots)  and results from the two fast-slow approximations (solid lines) for $f=0.8$ and three different ratios $\omega/\omega_o$. (a) $\omega/\omega_o=10$ where the fast boson approximation (blue lines) is valid. (b) $\omega/\omega_o=0.1$ where the fast pseudospin (red lines) is the valid one.  (c) and (d) correspond to the same $\omega/\omega_o=1.5$ ratio, where, according to Fig.\ref{fig:6}, neither   the fast boson [blue lines in panel (c)] nor the fast pseudospin [red lines in panel (d)] approximation  is valid. Insets show zooms to the low energy regions.   A  value $j=60$ was used.}
\label{fig:7}
\end{figure} 

In Fig.\ref{fig:7} three cases in the normal phase are shown. In panel (a) a ratio $\omega/\omega_o=10$ is used, where the fast boson approximation is expected to be valid. Since $\omega_B/\omega_F=16.67$, the interval of validity of the approximation extends to high energies, as can be seen
in the figure, where the Peres lattice is very well reproduced by the approximation. According to the fast boson approximation, the spectrum must be organized in finite bands labelled by the quantum number $n'$, extending from $E/(\omega_o j)=\omega/(\omega_o j) n'-1$ until $E/(\omega_o j)=\omega/(\omega_o j) n'+1$, exactly what is  observed in the numerical results. The end of the lowest band at $E/(\omega_o j)=1$ signals a so called Excited-State Quantum Phase Transition (ESQPT) \cite{Bas14}, where the whole pseudospin Bloch sphere becomes available. 

In panel (b) exact numerical and fast pseudospin approximation results are shown for a smaller ratio $\omega/\omega_o=0.1$, where the pseudospin to boson frequency ratio is $\omega_F/\omega_B=16.67$, and consequently the fast pseudospin approximation provides a good description of the exact results. This expectation is confirmed by the figure, where the Peres lattice is very well described by the approximated results. Observe that the energy scale is much smaller than the used in panel (a), that is because the density of states is much larger for this case and to achieve energies similar to those of panel (a), prohibitive matrix sizes would have to be used to obtain numerical exact results. Nevertheless, the number of lowest energy quantum states well  described by the approximation is similar to that of panel (a). According to the fast pseudospin approximation, the spectrum has to be organized in a finite number of infinite bands, the figure of panel (b) shows the first eleven of these bands, both in the approximated and numerical exact results.  

Finally in panel (c) and (d) the same  Peres lattice is shown but compared with results of the fast boson and fast pseudospin approximations respectively. The panels correspond to a ratio $\omega/\omega_o=1.5$ where, according to our previous analysis, neither one nor the other approximation are expected to give a good description of the exact results,  which is confirmed by the figures. Observe that in the case of the panel (c) a very small number of states  in the lowest energy band are described by the fast boson approximation. In the case of the fast pseudospin approximation not even this small number exists. This fact can be understood by looking at the pseudospin to boson ratios, according to the boson approximation this ratio  is $\omega_B/\omega_F=2.5$, whereas from the fast pseudospin one the ratio is very close to one $\omega_F/\omega_B=1.1$.  However, as discussed in section 3, in both cases the $\omega_{B,F}$ differ (for more than 5 \%) from the $\omega_\pm$ frequencies, and consequently both approximations have, as confirmed, to provide a very poor description of the exact results.

\begin{figure}
\begin{tabular}{cc}
(a)&(b)\\
\includegraphics[width=0.47\textwidth]{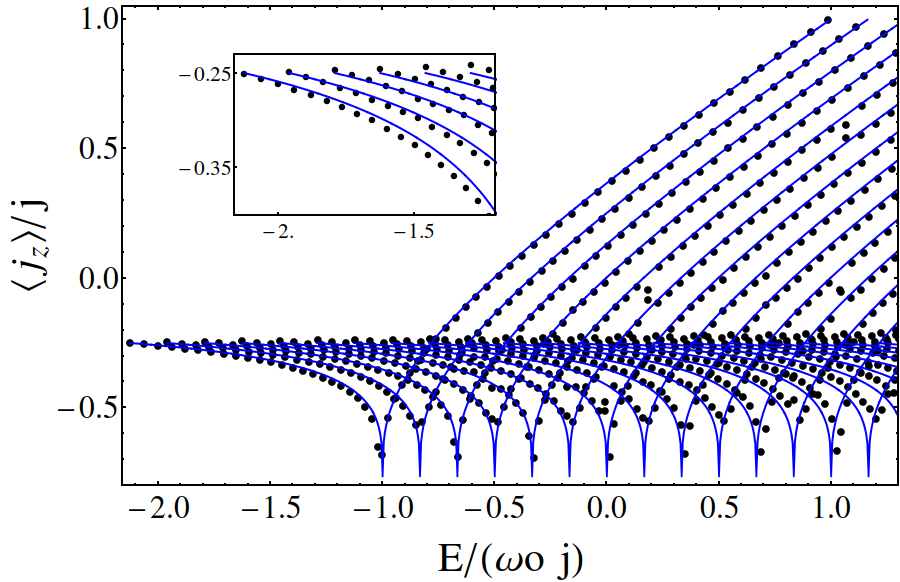}&\includegraphics[width=0.48\textwidth]{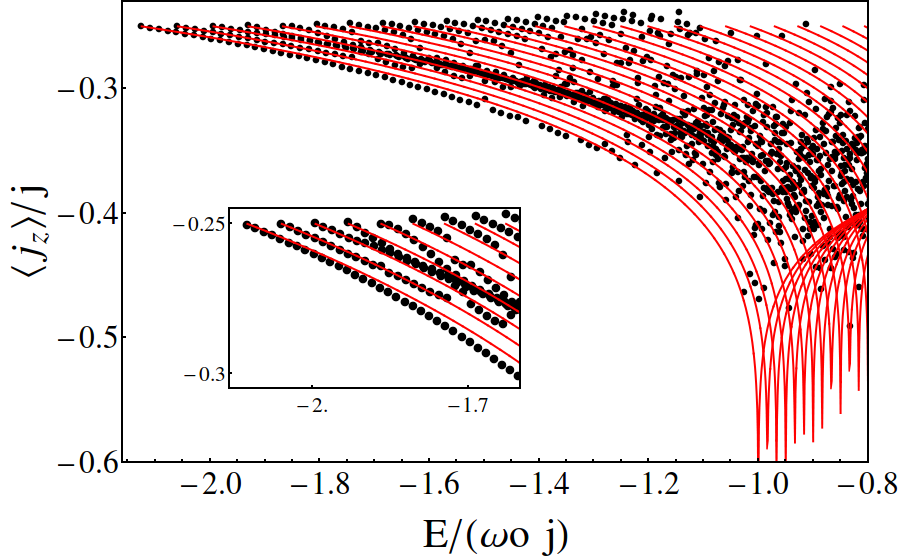}\\
(c)&(d)\\
\includegraphics[width=0.47\textwidth]{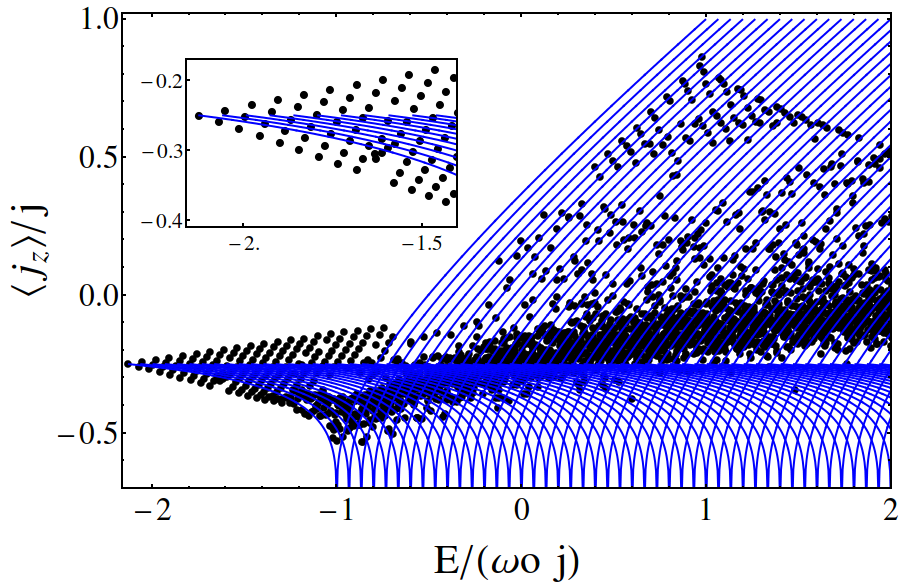}&\includegraphics[width=0.47\textwidth]{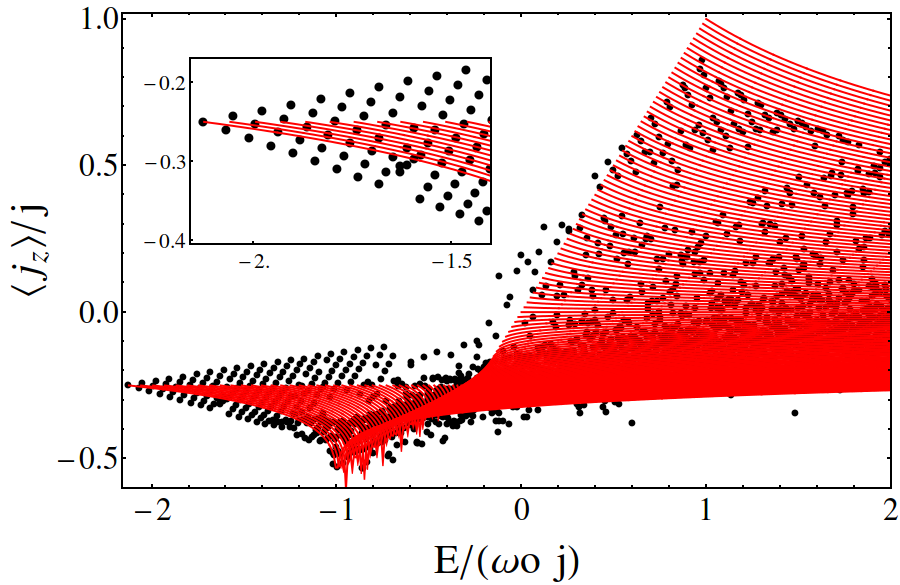}\\
\end{tabular}
\caption{ Similar to Fig.\ref{fig:7} but for $f=2$ and $\omega/\omega_o=10$ (a), $\omega/\omega_o=1$ (b) and $\omega/\omega_o=4$ (c) and (d).}
\label{fig:8}
\end{figure} 

In Fig.\ref{fig:8} three cases in the superradiant phase with $f=2$ are shown. In panel (a) a ratio $\omega/\omega_0=10$ is employed, which corresponds to a ratio $\omega_B/\omega_F=5.77$. In  accord to our previous analysis, the fast boson approximation (blue lines), describes very well the numerical exact results (dots) until very large excitation energies. Even if some numerical discrepancies can be seen (attributable to the fact that the ratio $\omega_B/\omega_F=5.77$ is not much larger than $1$), the overall structure of the Peres lattice is completely reproduced by the fast boson approximation. The finite bands (labelled by quantum number $n'$)  are clearly distinguishable in the Peres lattice, and  the two lowest bands can be seen completely (more complete bands are hidden because of the plot range used). According to the approximation, the lowest band ($n'=0$) must begin at energy $E/(\omega_o j)=-\frac{1}{2}(f^2+f^-2)$ and ends at $E/(\omega_o j)=1$, the energy of the ESQPT where the whole Bloch sphere becomes available. These features are clearly seen in the exact numerical results. 
It is known \cite{Per11,Bran13,Bas14} that in the superradiant phase, a second ESQPT appears at energy $E/(\omega_o j)=-1$. A singular behaviour of the lowest band is observed exactly at this
energy, and observe that every band present a similar singular behaviour at energies $E/(\omega_o j)=-1+\omega/(\omega_o j) n'$, which are associated to a logarithmic divergence in the Density of
States $\nu^a(E,n')$ of the one degree of freedom effective Hamiltonian $H_{n'}\left(\vec{j}\,\right)$, in agreement with the general theory of ESQPTs of Ref.\cite{Stransky15}. Notwithstanding, it is worth remarking that the ESQPTs already observed in the Dicke model do not show divergences in the
Density of States, but in its first  derivative \cite{Per11,Bran13,Bas14}. This is because these ESQPTs are observed without considering the invariant subspaces resulting from the BOA adiabatic invariants, which is the relevant  physical situation within the chaotic region, where the second adiabatic invariant does not exist anymore and thus the dynamics spreads over the two degrees
of freedom of the Hamiltonian. But if the BOA is valid until the energy of the ESQPTs [as the case shown in panel (a) of Fig.\ref{fig:8}], therefore, our results show that the critical behavior in the spectrum of the Dicke model dramatically changes. From a divergent Densities of States for each invariant band in the region in which the BOA is valid, to a continuous Density of States (with divergent derivative) in the region where is not.

The case shown in panel (b) for a ratio $\omega/\omega_0=1$, corresponds to a ratio $\omega_F/\omega_B=4.13$, where  the fast pseudospin approximation is valid, at least in a certain energy interval  above the ground-state energy. As the ratio $\omega_F/\omega_B=4.13$  is not much larger than $1$, is expected that this energy interval does not extend too far from the ground state. According to the fast pseudospin approximation, the spectrum has to be organized in a finite number ($2j+1$) of  bands labelled by the quantum number $m'$, the Peres lattice shows only a small fraction of them (about 16), and fewer are correctly reproduced by the fast pseudospin approximation (around the first 6). The number of lowest states correctly reproduced by the fast pseudospin approximation decreases as we look at more excited bands. The fast pseudospin approximation provides a qualitative, and sometimes
quantitative, description of the regular part of  the Peres lattice, until the onset of chaos in the model. It is worth mentioning that the fast pseudospin approximation gives a more accurate and extended description of the exact spectrum than the Holstein-Primakoff approach \cite{Relano16}, the only analytic approximation to the low energy region of the Dicke model known so far.   

In panels (c) and (d), corresponding to $\omega/\omega_o=4$ and $\omega_F/\omega_B\sim 1$, the numerical exact results confirm what is expected from the analysis of section 3. None of the fast-slow approximations give a good description of the Peres lattice, not even in the low energy part. It interesting to note that the Peres lattice of this case is completely regular in the low energy sector, remaining as a challenge to obtain an analytical approximation to the Dicke model able to describe it.  

\begin{figure}
\begin{tabular}{cc}
(a)&(b)\\
\includegraphics[width=0.48\textwidth]{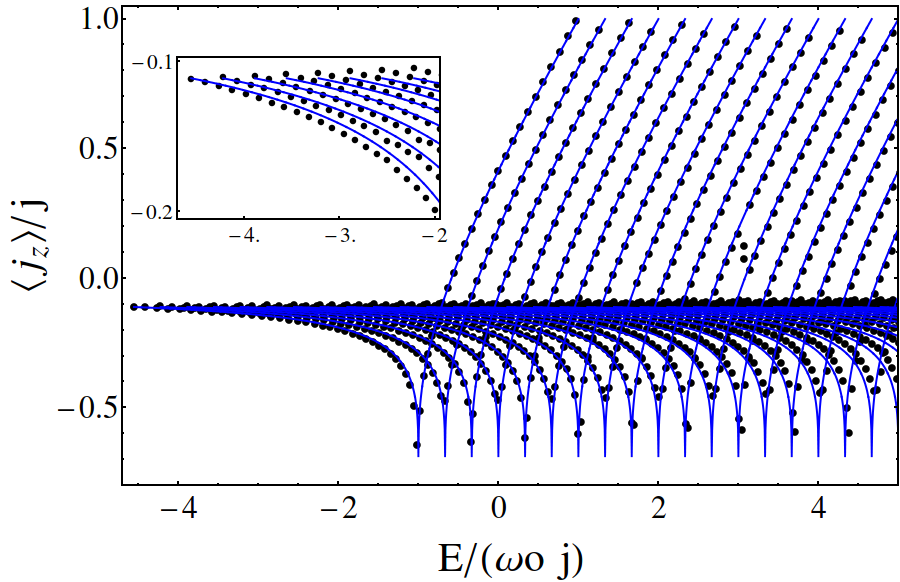}&\includegraphics[width=0.48\textwidth]{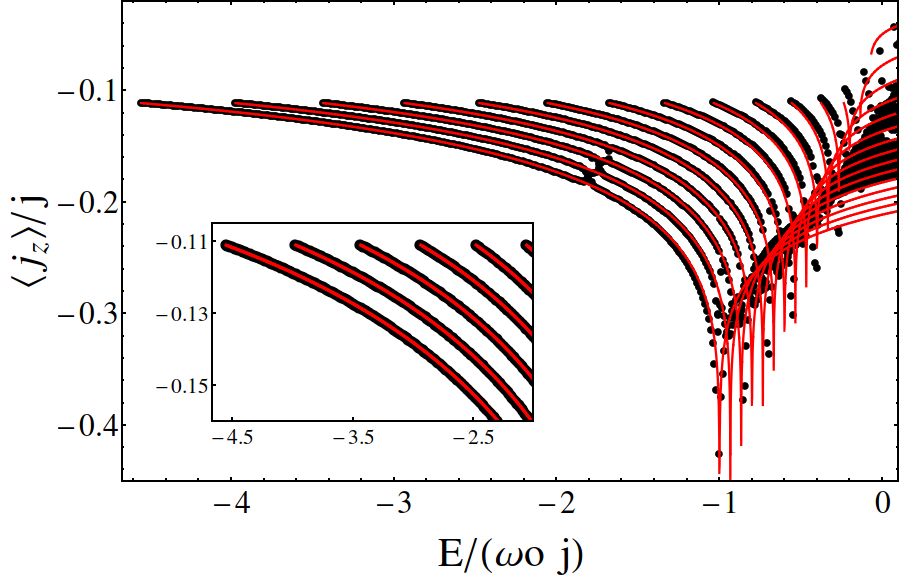}\\
(c)&(d)\\ 
\includegraphics[width=0.48\textwidth]{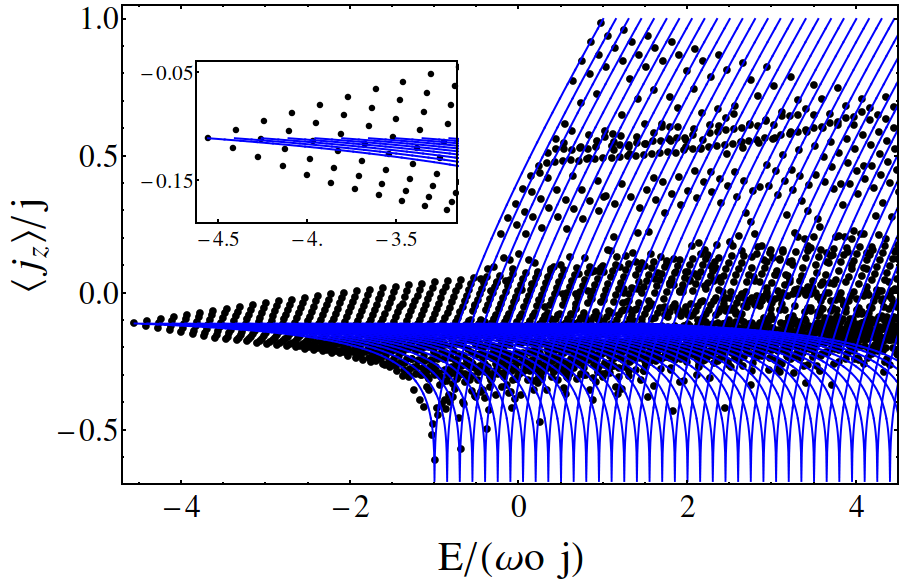}&\includegraphics[width=0.48\textwidth]{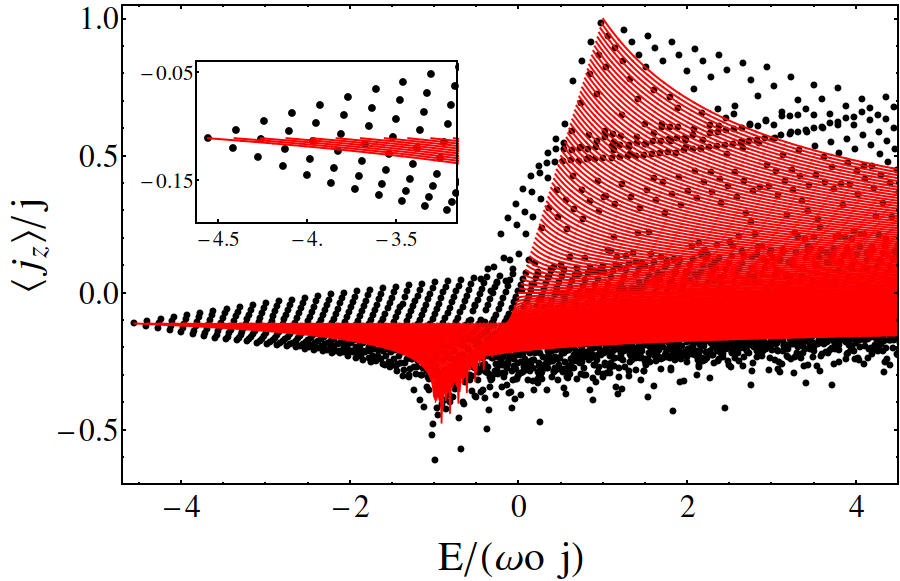}\\
\end{tabular}
\caption{Similar to Fig.\ref{fig:7} but for $f=3$ and $\omega/\omega_o=20$ (a), $\omega/\omega_o=0.2$ (b) and $\omega/\omega_o=9$ (c) and (d). In panel (b) a $j=15$ value was used.}
\label{fig:9}
\end{figure} 

In Fig.\ref{fig:9} a large coupling $f=3$ in the superradiant phase is considered. In panel (a) a ratio $\omega/\omega_o=20$ is used, in the region of applicability of the fast boson approximation. Even if $\omega/\omega_o$ is much larger than $1$, the boson to pseudospin ratio is not so large  $\omega_B/\omega_F=2.24$. The exact Peres lattice is globally very well described by the approximation, though numerical discrepancies can be observed in the low energy region. The expected finite bands of the approximation are clearly seen in the numerical results. Every band exhibits its own ESQPT at $E/(\omega_o j)=-1+\omega/(\omega_o j) n'$ and ends at energies close to $E/(\omega_o j)=1+\omega/(\omega_o j) n$. 

In panel (b), fast pseudospin approximation results are shown together with exact numerical results for $\omega/\omega_o=0.2$, which gives   $\omega_F/\omega_B=45.28$.  Since this particular case is numerically more challenging, a much more modest system size was employed ($j=15$), even so the number of sates considered to make the comparison is similar to the other plots where $j=60$ was used. The energy range considered includes the energy of the ESQPT ($E/(\omega_o j)=-1$), and almost all the exact states below this energy are very well reproduced by the approximation. The approximation seems to work very well until the onset of chaos in the system, which appears at large excitation energies, larger for more excited bands (larger $m'$).

In the low (c) and (d) panels, the same Peres lattice for $\omega/\omega_o=9$ is shown, in panel (c) is compared with the results of the fast boson approximation and in panel (d) with those of the fast pseudospin one. Except for the ground-state, as expected from the analysis of section 3, the exact Peres lattice is not even closely reproduced by the results of any approximation.  Although in this case no approximation gives a close description of the exact results, as in the previous figures, the low energy part of the Peres lattice is regular. As mentioned, it would be interesting to find an approximation able to describe this regular region. 

\section{Conserved quantities}

Associated to one and other fast-slow approximation, adiabatic invariants can be identified. These adiabatic invariants are, to the extent that one or other approximation is valid, approximate integrals of motion. In this section, these adiabatic invariants are given and discussed. Numerical results are presented  to illustrate that these invariants are effectively approximated integrals of motion. Numerical results are presented  both in the quantum and classical version of the Dicke model.

\subsection{Adiabatic invariant in the fast pseudospin approximation}

In the case of the fast pseudospin approximation, the adiabatic invariant is the projection of the  pseudospin along the precession axis, which, in turn, is a function of the  $q$ coordinate of the slow bosonic variables
\begin{equation}
J_{z'}=\frac{\omega_o}{\sqrt{\omega_o^2+\left(\frac{2\gamma}{\sqrt{j}}q\right)^2}}J_z+\frac{\frac{2\gamma}{\sqrt{j}}q}{\sqrt{\omega_o^2+\left(\frac{2\gamma}{\sqrt{j}}q\right)^2}}J_x.
\end{equation}
In the classical version of the model, the previous expression gives the approximate integral of motion by considering the pseudospin components as classical variables, $J_i\rightarrow j_i$, of a vector of magnitude $|\vec{j}|=j$. In the quantum case, the corresponding quantum observable is obtained by substituting the $q$ variables by creation and annihilation operators [$q\rightarrow (a+a^\dagger)/\sqrt{2}$]
\begin{equation}
J_{z'}=\frac{\omega_o}{\sqrt{\omega_o^2+\left(\frac{\sqrt{2}\gamma}{\sqrt{j}}(a+a^\dagger)\right)^2}}J_z+\frac{\frac{\sqrt{2}\gamma}{\sqrt{j}}(a+a^\dagger)}{\sqrt{\omega_o^2+\left(\frac{\sqrt{2}\gamma}{\sqrt{j}}(a+a^\dagger)\right)^2}}J_x.
\end{equation}

\begin{figure}
\begin{tabular}{lr}
\includegraphics[width=0.7\textwidth]{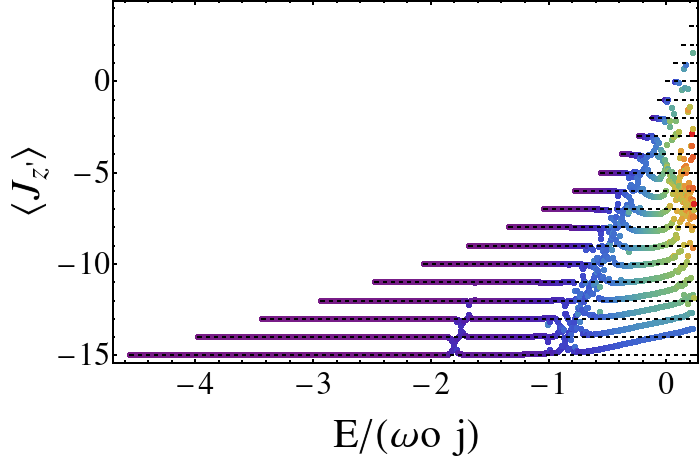}& \includegraphics[width=0.13\textwidth]{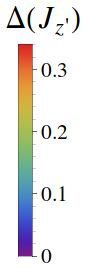}
\end{tabular}
\caption{Peres lattice for  operator $J_{z'}$ for $f=3$, and $\omega/\omega_o=0.2$. The colors indicate the uncertainty $\Delta J_{z'}$ evaluated in the respective Hamiltonian eigenstate. Horizontal dashed lines indicate the values $m'=-j,-j+1,...$, the lines begin  at the energies predicted by the fast pseudospin approximation. }
\label{fig:10}
\end{figure}

In Fig.\ref{fig:10}  full quantum results are presented.  The Peres lattice of  $J_{z'}$ is shown for $f=\gamma/\gamma_c=3$ and a ratio $\omega/\omega_o=0.2$ [the same parameters used in panel (b) of Fig.\ref{fig:9}]. The results show that, in the low energy region where the approximation is valid,  the Hamiltonian eigenstates are organized in horizontal bands associated each with the quantum number $m'$ (horizontal dashed lines). The calculation of the uncertainty $\Delta J_{z'}$ gives values very close to zero for the lowest energy eigenstates of each band, showing that these Hamiltonian eigenstates are  simultaneously very approximated eigenstates of operator $J_{z'}$.  

\begin{figure}
\begin{tabular}{cc}
(a)&(b)\\
\includegraphics[width=0.4\textwidth]{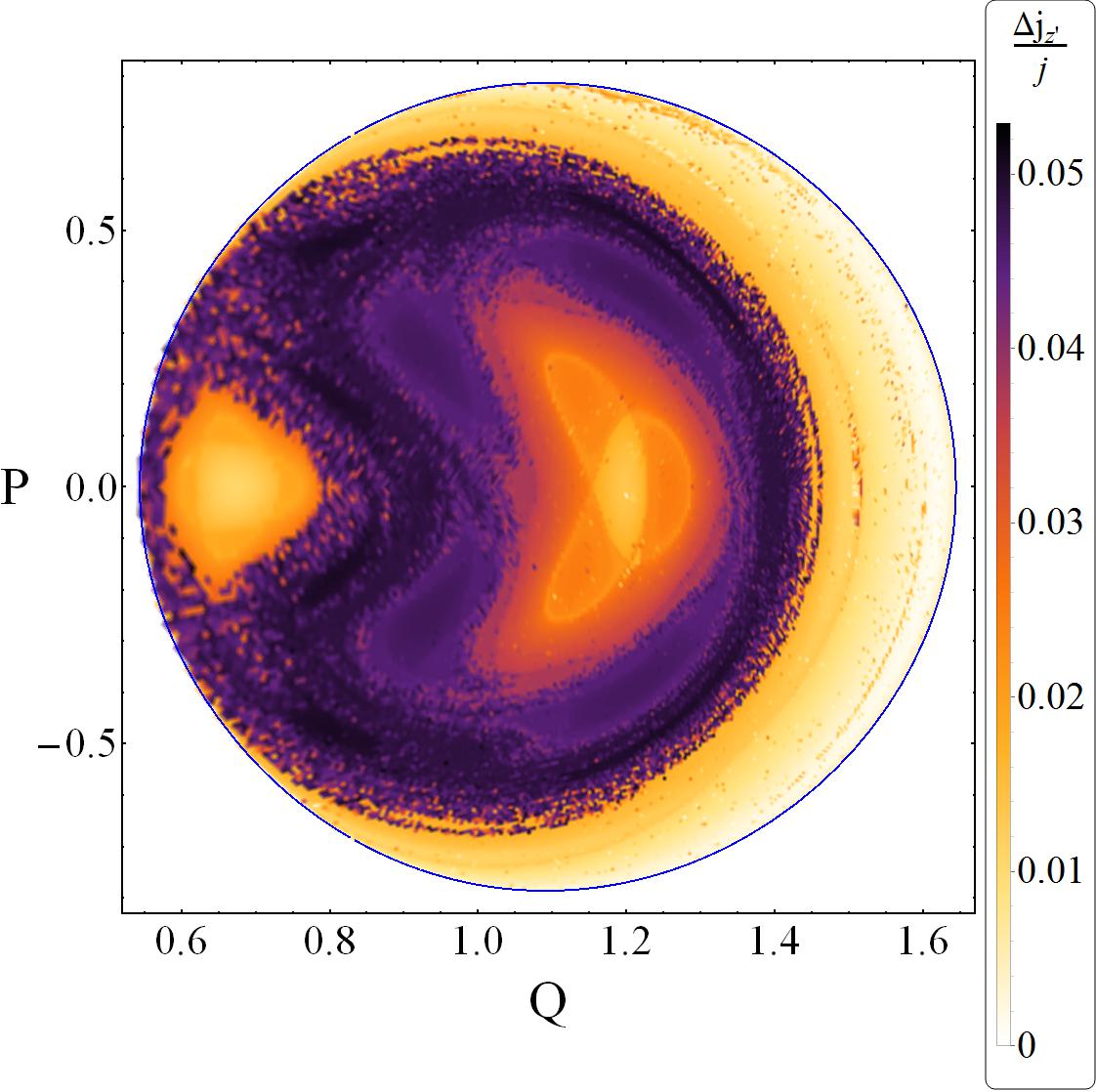} &\includegraphics[width=0.4\textwidth]{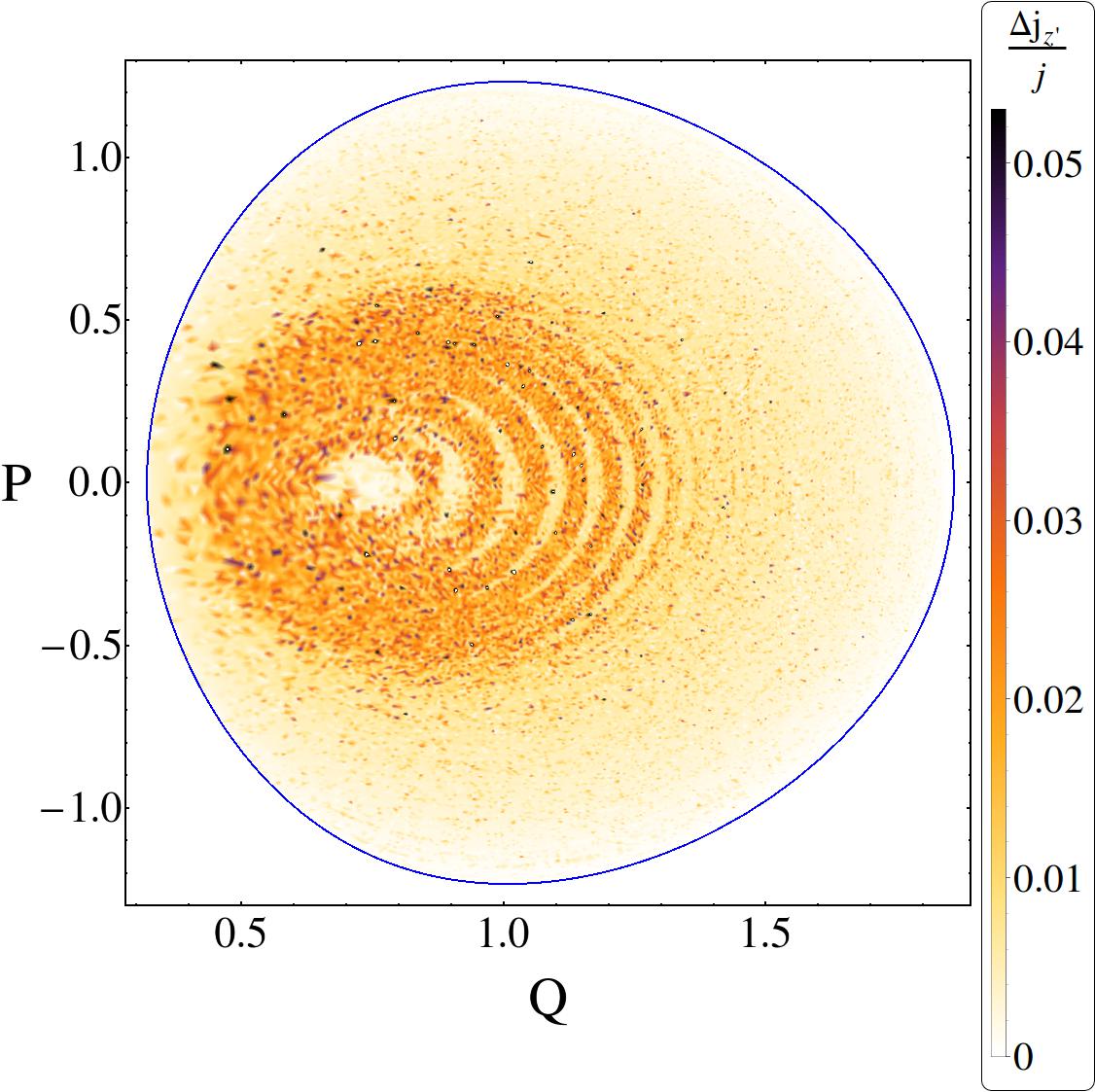} \\
(c)&(d)\\
\includegraphics[width=0.33\textwidth]{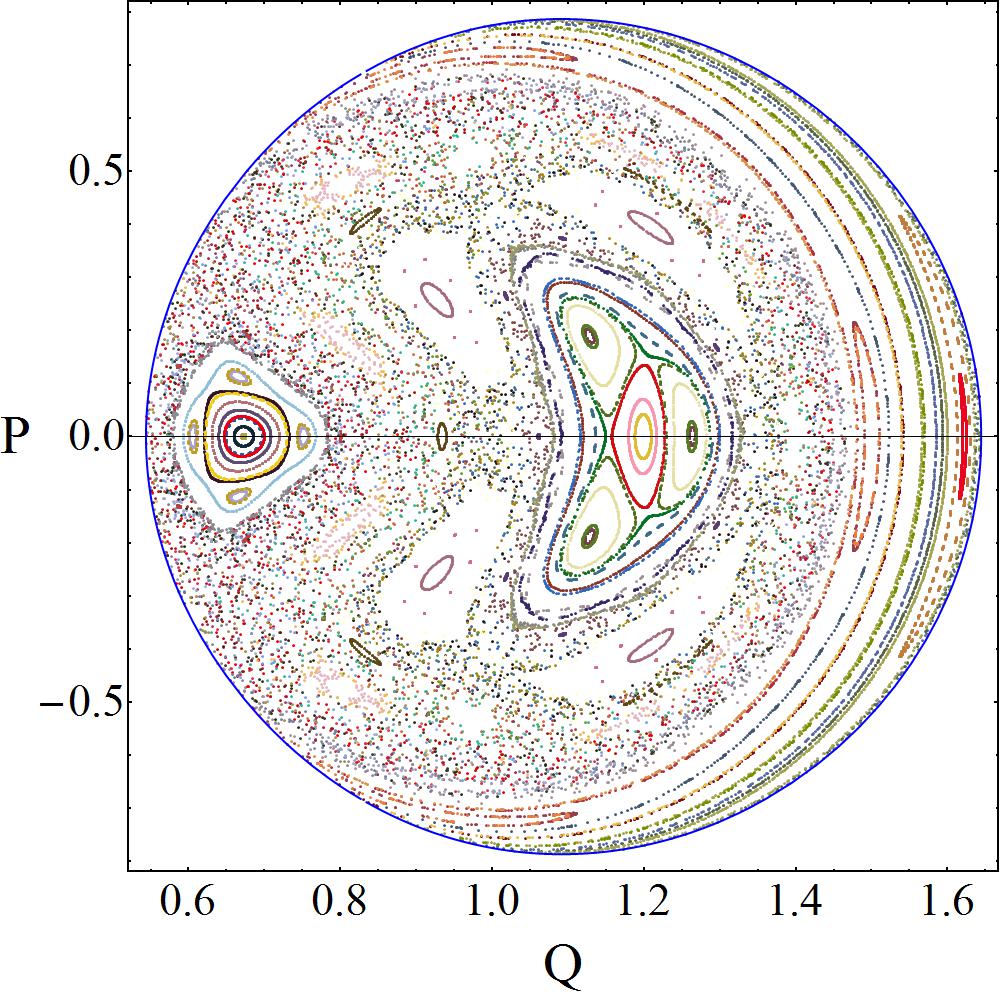}&\includegraphics[width=0.33\textwidth]{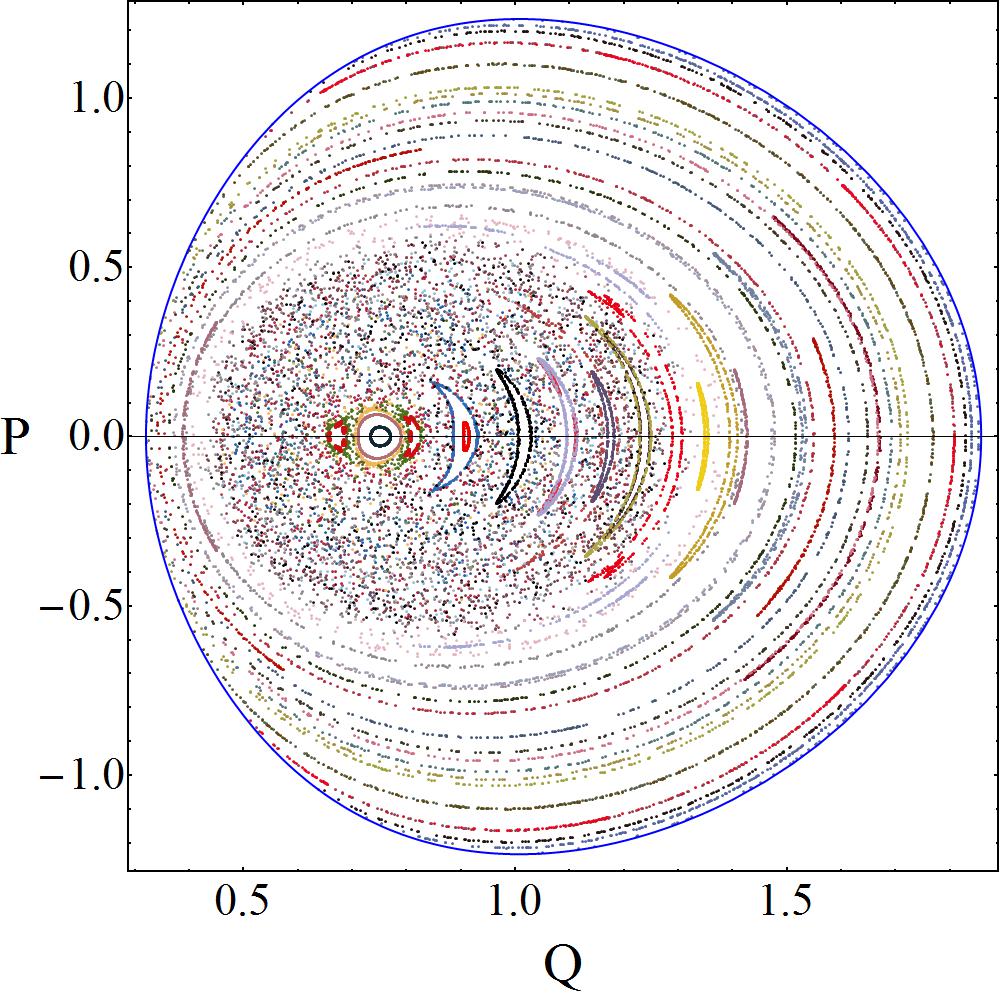}\\
\end{tabular}
\caption{Temporal variances $\Delta j_z'/j$ of the classical Dicke model for a wide sample of initial conditions in the energy surface  $E/(\omega_o j)=-1.4$. Panel (a) is   for $f=2$ with $\omega/\omega_o=1$ and panel (b) for $f=3$ with $\omega/\omega_o=0.2$ (b). Panels (c) and (d) show, respectively,  Poincar\'e sections for the same parameters and energy.}
\label{fig:11}
\end{figure}

In Fig.\ref{fig:11}, similar results are presented, but for the classical version of the model (see \ref{appendixC} for the definition of the classical  Hamiltonian employed). In panels (a) and (b)  the variance of temporal averages of the  classical variable $j_{z'}$
$$
\Delta j_{z'}=\sqrt{\frac{1}{T}\int_0^T (j_{z'}(t))^2 dt-\left(\frac{1}{T}\int_0^T j_{z'}(t)dt\right)^2 },
$$
is  calculated integrating the classical trajectories between the times $0$ to $T$, for a a wide sample of initial conditions in a surface of constant energy [$E/(\omega_o j)=-1.4$], with the same sets of parameters as panels (b) of Figs.\ref{fig:8} and \ref{fig:9} respectively, $f=2$ and $\omega/\omega_o=1$ in panel (a) and  $f=3$ and $\omega/\omega_o=0.2$ in panel (b). The variance of $j_{z'}$ takes values very close to zero (indicating at what extent  $j_{z'}$ is an approximated   constant of motion)  for initial conditions in the regular part of the phase space. Panels (c) and (d) display the corresponding Poincar\'e sections, allowing to identify regular and chaotic regions. For initial conditions in the chaotic regions of the phase space, the variance of $j_{z'}$ takes large values. These results suggest that the onset of chaos in the model is intimately related with the breaking of the fast pseudospin approximation, where the dynamical variable  $j_{z'}$ ceases to be an approximated integral of motion.

\begin{figure}
\begin{center}
\includegraphics[width=0.7\textwidth]{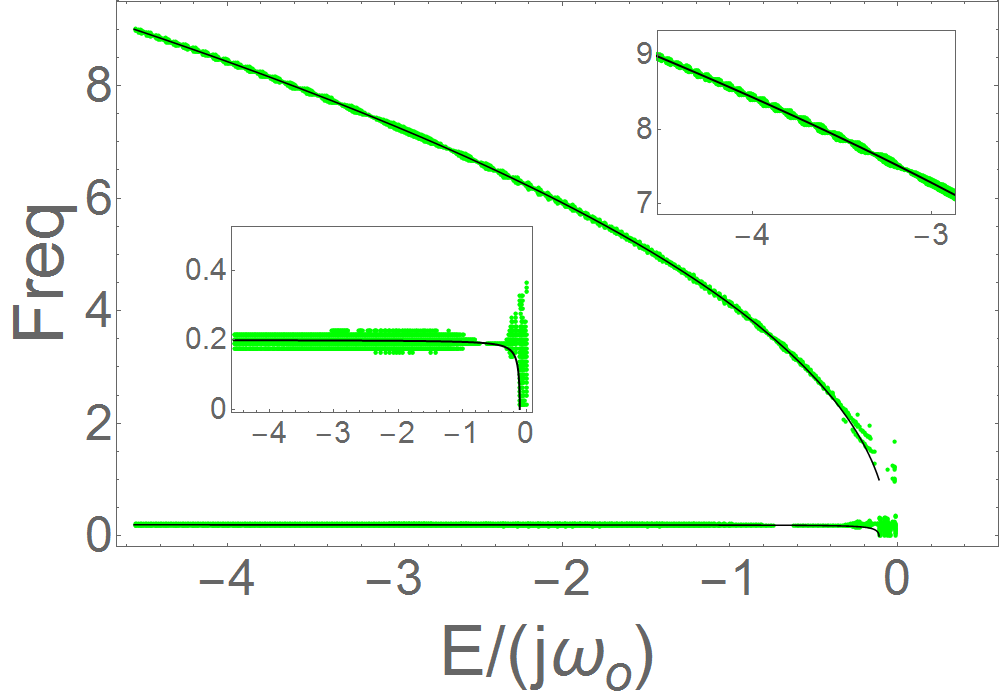}
\end{center}
\caption{Light dots (green online) depict the principal frequencies of the exact dynamical classical variable $q$, as a function of energy, for initial conditions corresponding to the minimal energy configurations of the effective potential $V_{m'}(q)$. Solid black lines are the analytic pseudospin and bosonic frequencies obtained from the fast pseudospin approximation Eqs. (\ref{eq:frecBos}) and (\ref{eq:frecPseu}). Insets show a closer view of the fast and slow frequencies. The system considered was $\omega/\omega_o=0.2$ with  $f=3$.}
\label{fig:12}
\end{figure}

A complementary result showing the ability of the fast pseudospin approximation to explain and reproduce the regular energy regime  of the classical Dicke model, is shown in Fig.\ref{fig:12} for the case $\omega/\omega_o=0.2$ with  $f=3$. We select as initial conditions those corresponding to the minimal configuration energy for the effective  potentials $V_{m'}(q)$, for different values of $m'\in[-j,j]$ (in the classical version $m'$ is a continuous variable). We integrate the classical equations of motion, obtaining the classical trajectories for these initial conditions. A Fourier analysis of a given dynamical variable (in this case $q$) allows to extract its fundamental frequencies. The most relevant classical
frequencies are compared with those obtained from the fast pseudospin approximation given in Eqs. (\ref{eq:frecBos}) and (\ref{eq:frecPseu}). The agreement between the set of classical frequencies 
and their analytic approximated estimation is very good, until the onset of chaos in the system, which occurs at energy $E/(\omega_o j)\sim 0$ for the band head states (lowest energy states of each $m'$),  as can be seen in panel (b) of Fig.\ref{fig:9}.    
 
\subsection{Adiabatic invariant in the fast boson approximation}

In the case of the fast boson approximation, the adiabatic invariant is the number of quanta of the shifted operator $b^\dagger b$, which written in terms of the boson and spin variables is
\begin{equation}
b^\dagger b=a^\dagger a+\frac{2\gamma^2}{j\omega^2}J_x^2+\sqrt{\frac{2}{j}}\frac{\gamma}{\omega}(a+a^\dagger)J_x.
\end{equation}

\begin{figure}
\begin{tabular}{cc}
\includegraphics[width=0.7\textwidth]{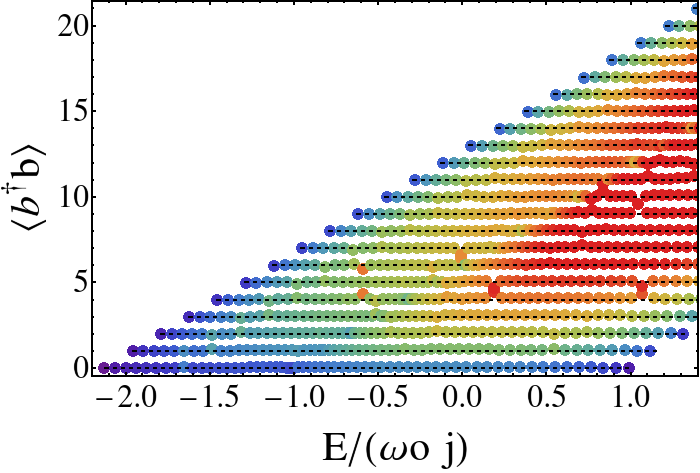}&\includegraphics[width=0.15\textwidth]{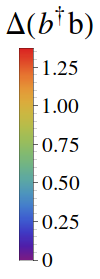}
\end{tabular}
\caption{Peres lattice of operator $b^\dagger b$ for $f=2$ and $\omega/\omega_o=10$. The point colors indicate the uncertainty of the respective Hamiltonian eigenstate. Horizontal dashed lines indicate the values $n'=0,1,...$, the lines begin and finish at the energies predicted by the fast boson approximation.}
\label{fig:13}
\end{figure}

In Fig.\ref{fig:13}, the Peres lattice of operator $b^\dagger b$ is shown for the case $f=2$, $\omega/\omega_o=10$ [the same parameters used in panel (a) of Fig.\ref{fig:8}]. It can be seen that the finite bands are associated  to the quantum number $n'=0,1,...$ (dashed lines). The colors of the points in the Peres lattice indicate the uncertainty  $\Delta b^\dagger b$. Observe that the lowest states of every band and all the members of the lowest band ($n'=0$) have very low uncertainty, which indicates that these Hamiltonian eigenstates are, in some extent, also simultaneous eigenstates of operator $b^\dagger b$. 

\subsection{Requantization}

Finally, we discuss the full quantum treatment that can be performed from the fast-slow approximations. In the case of the fast pseudospin approximation, once the pseudospin part has been diagonalized by considering a rotation around the $y$ axis, we are left with the Hamiltonian $H_{m'}=\frac{\omega}{2}p^2+V_{m'}(q)$, Eq.(\ref{Hmp}), for each $m'$. This Hamiltonian can be quantized in the standard way to obtain a Schr\"odinger equation
$$
\left(-\frac{\omega}{2}\partial_q^2+ V_{m'}(q)\right) \Psi_E(q)=E\Psi_E(q),
$$
or alternatively, if we are interested only in the energy spectrum,  a standard Bohr-Sommerfeld quantization can be employed 
\begin{equation} \oint_{C_{m'}} p dq=2\pi n,
\end{equation} 
where $C_{m'}$ is a closed path in the classical phase space associated to the Hamiltonian $H_{m'}$. In any case the result is a spectrum associated to the band $m'$ of the complete spectrum. 

\begin{figure}
\begin{center}
\includegraphics[width=0.5\textwidth]{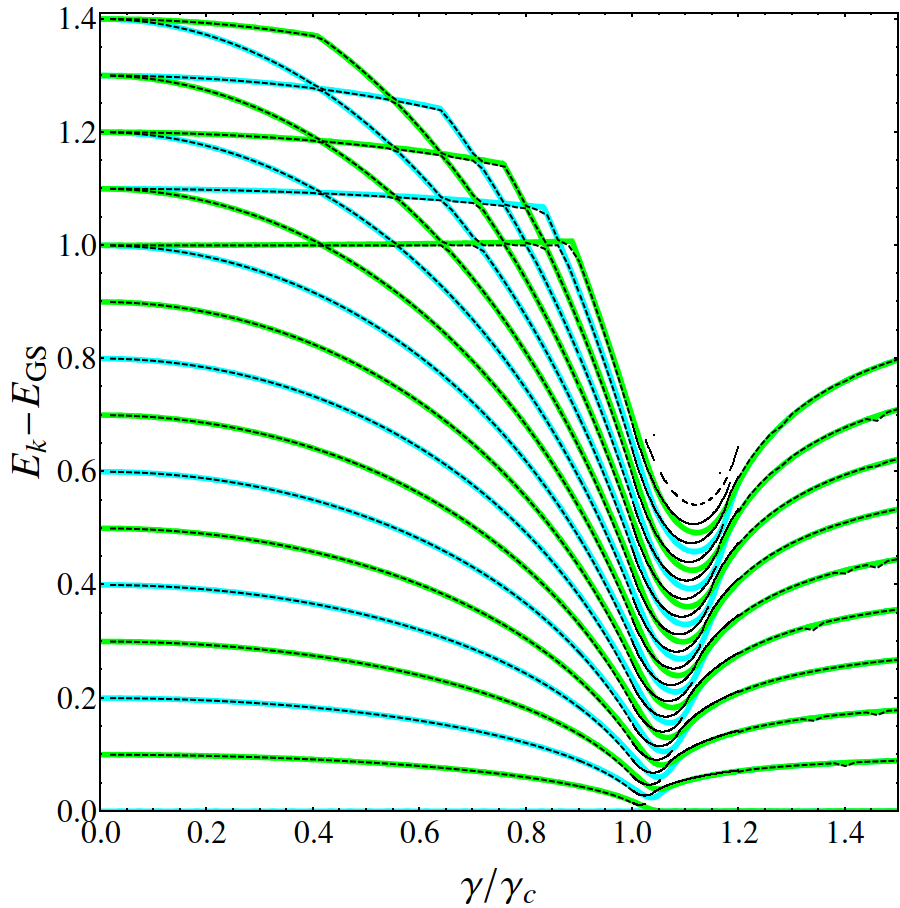}\ \ \includegraphics[width=0.25\textwidth]{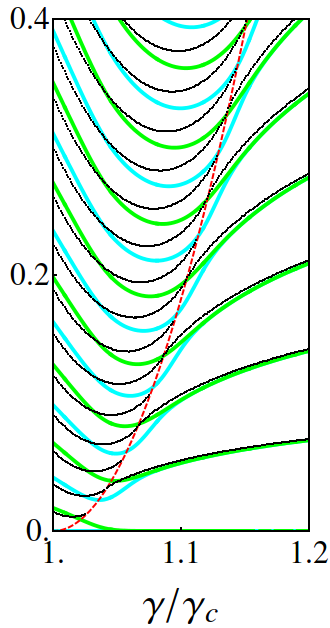}
\end{center}
\caption{Excitation energy  of the first 20 states of the Dicke model as a function of the coupling in the  case $\omega/\omega_o=0.1$ with $j=10$.  The exact spectrum is shown by solid light lines, the blue and green  (color online) lines are for the positive and negative parity sector respectively. The dashed black lines are the results of the fast pseudospin BOA using the  Bohr-Sommerfeld quantization rule. Right panel shows a closer view of the region around the critical coupling, the red dashed line indicates the critical energy of the ESQPT.}
\label{fig:14}
\end{figure}

Fig.\ref{fig:14} shows the results of the latter approach, compared with numerical exact results for couplings ranging from 0 until the superradiant phase, in the case $\omega/\omega_o=0.1$. Only the lowest 20 energy states are plotted. The five states with higher energies look broken because they have avoiding crossing with other higher states not shown. The approximate results reproduce remarkably  well the exact spectrum, except in the critical ESQPT energy region of the superradiant phase, where the tunneling effects, not considered by the  Bohr-Sommerfeld quantization rule, become relevant. In Ref.\cite{Relano16}, results using  the same  approach are given and it is shown that the approximate and exact spectrum get closer as the number of two-level systems $j$ increases. 

For the fast-boson approach, the procedure is similar, after diagonalizing the fast boson variable by using the shift transformation, we are left, for every quantum number $n'$, with an effective Hamiltonian for the pseudospin variables (\ref{Hnp}). By considering the pseudospin components as $SU(2)$ quantum operators, the resulting Hamiltonian is a particular version of the Lipkin-Meshkov-Glick model that  can be numerically diagonalized, giving the spectrum associated to band $n'$. It is interesting to note that the same LMG model is obtained for each band $n'$, and the only difference between the bands' spectra is an additive simple constant $\omega n'$. 

\begin{figure}
\begin{center}
\includegraphics[width=0.7\textwidth]{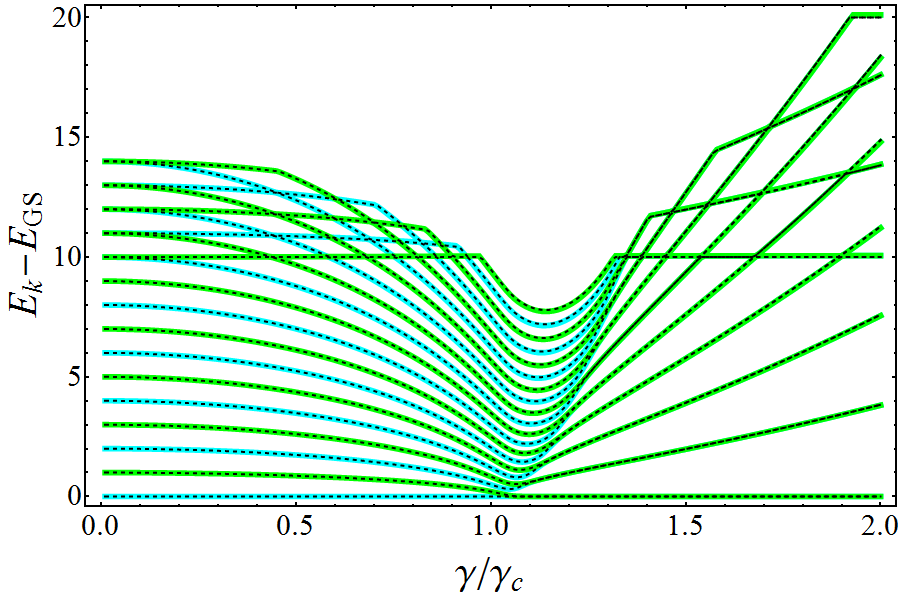}
\end{center}
\caption{Similar to Fig.\ref{fig:14} but for $\omega/\omega_o=10$ with $j=60$, and the approximate results are obtained from the fast boson approximation.}
\label{fig:15}
\end{figure}

In Fig.\ref{fig:15} we compare the approximate spectrum obtained from the previous  approach with the exact one obtained numerically, we use a ratio $\omega/\omega_o=10$ and couplings from $0$ until $\gamma=2\gamma_c$. Similar to the fast pseudospin case, the agreement is very good. 

\section{Conclusions}

A complete survey of the Born-Oppenheimer approximation (BOA) applied to the Dicke model has been presented. The study includes both the case when the fast variable is the pseudospin one and the case when the fast variable is the bosonic one. The ranges of validity of both versions of the BOA were unveiled. It was found that in the normal phase a simple criterion is  $\omega/\omega_o>>1$ for the fast boson BOA and $\omega/\omega_o<<1$ for the fast pseudospin one, but in the superradiant phase, as a consequence of the change in the fundamental effective excitations, the criterion changes to $\omega/\omega_o>>f^2$ and $\omega/\omega_o<<f^2$ for the fast boson and fast pseudospin BOA's respectively ($f=\gamma/\gamma_c$ with $\gamma_c=\sqrt{\omega \omega_o}/2$).

More restrictive and detailed criteria were obtained by comparing the low energy frequencies coming from the BOA's with those obtained from a quadratic approximation around the minimal energy configuration. Ample numerical tests were presented to show the effectiveness of these criteria and the ability of the BOA's to reproduce the exact results of the Dicke model, both in its classical and quantum version. It was found that in the BOA's reproduce very well the exact results in the normal and superradiant phase, from the ground state until the energy where chaos appears in the model. The approximations explain in a clear way the appearance of independent energy bands in the regular part of the spectrum in a wide region of the model parameter space. Quantitative discrepancies are found, but they are reduced for larger ratios $\omega_B/\omega_F$ in the case of the fast boson BOA and $\omega_F/\omega_B$ in the case of the fast pseudospin one. 

The adiabatic invariants associated to the two different BOA's were explicitly identified and they were found to give very approximate Hamiltonian commuting operators in the quantum version of the model and very approximate conserved quantities in its classical one. Our results show that the BOA is very efficient to explore and explain the phenomenology observed in the regular regime of the Dicke model, up to energies where chaos appears and no other approximation known so far is valid.

Besides the understanding of the spectrum in a wide region above the ground-state, these results provide a valuable tool to study dynamics in the Dicke model. In particular, we expect that the second adiabatic invariant determines the non-equilibrium dynamics and the further thermalization after a quench, leading the system to a generalized Gibbs ensemble, instead of a standard Gibbs ensemble, within the region in which either the fast boson or the fast pseudospin approximation is valid. Also, as we have foreseen in this work, the properties of the ESQPTs in this model are delimited by the existence (or not) of the second adiabatic invariant. Within the region in which the BOA is valid, the critical behavior is the corresponding to a system of just one degree of freedom, showing a logarithmic divergence in the density of states. It is worth to note that the same kind of critical behavior has been recently reported in the Rabi model, in the limit in which the atomic frequency is much larger than the frequency of the field \cite{Puebla}, that is, in a region in which the fast pseudospin approximation is expected to hold.

\section*{Acknowledgments} A. R. is supported by Spanish Grants No. FIS2012-35316 and FIS2015-63770-P (MINECO/FEDER), M. A. B-M, S.L-H, B. LC, J. C-C and J.G. H acknowledge financial support from Mexican CONACyT project CB166302 and DGAPA-UNAM project IN109417. S.L-H. acknowledges financial support from Mexican CONACyT project CB2015-01/255702 and from CONACyT fellowship program for sabbatical leaves.


\appendix

\section{Low energy frequencies of slow variables}
\label{appendixA}

In this appendix we derive the lowest energy  frequencies of the slow variables for the two BOA's, by expanding the effective Hamiltonians  around the minimal energy configurations until quadratic terms, $H=\frac{1}{2}A p^2+\frac{1}{2}B q^2$, and using the very well known result for the frequency of quadratic Hamiltonians $\omega=\sqrt{AB}$.  

\subsection{Fast pseudospin: low energy boson frequencies for each $m'$}

 We expand the potential $V_{m'}(q)$ appearing in the effective Hamiltonian (\ref{Hmp}), around the minimal energy configuration  given in Eq.(\ref{eq:qmin}). When  the potential is a double well ($-j\leq m'\leq -j/f^2$) this expansion gives
$$
H_{m'}=E_{min}^{m'}+\frac{\omega}{2} p^2 +\frac{\omega}{2}\left(1-\left(\frac{j}{m'}\right)^2\frac{1}{ f^4}\right)(q-q_{min}^{m'})^2+ ... ,
$$
therefore the boson frequency is
$$
\omega_B(E_{min}^{m'},m')=\sqrt{\omega^2\left(1-\left(\frac{j}{m'}\right)^2\frac{1}{ f^4}\right)}=\omega\sqrt{\frac{f^4-\left(\frac{j}{m'}\right)^2}{f^4}}.
$$
When the potential has a single minimum at $q=0$, the expansion of the potential around it, yields
$$
H_{m'}=\omega_o m'+ \frac{\omega}{2} p^2+\frac{\omega}{2}\left(1+\frac{m'}{j}f^2\right)q^2+...,
$$
from where the boson frequency can be easily obtained
$$
\omega_B(E_{min}^{m'},m')=\sqrt{\omega^2\left(1+\frac{m'}{j}f^2\right)}=\omega\sqrt{1+\frac{m'}{j}f^2}.
$$

\subsection{Fast boson: low energy  pseudospin frequency}

To find  the slow pseudospin frequencies, we express the effective Hamiltonian (\ref{Hnp}) in terms of the canonical variables shown in Eq.(\ref{eq:cano})
$$
H_{n'}(Q,P)=\omega n'-\omega_o j+\frac{\omega_o j}{2}\left[\frac{P^2}{j^2}+Q^2-\frac{f^2 Q^2}{4}\left(4- \frac{P^2}{j^2}-Q^2\right)\right].
$$
Then, we expand this Hamiltonian around the minimal energy configuration given in Eq.(\ref{eq:minconFB}). For $\gamma<\gamma_c$, the minimal energy configuration in terms of the canonical variables is $(Q_{min},P_{min})=(0,0)$. The expansion of the  Hamiltonian until quadratic terms  is
$$
H_{n'}(Q,P)=\omega n'-\omega_o j+\frac{1}{2}\frac{\omega_o}{j}P^2+\frac{1}{2}\omega_o j(1-f^2)Q^2+...
$$
From this expression we obtain the pseudospin frequency
$$
\omega_F(E_{min}^{n'})=\sqrt{(\omega_o/j)\omega_o j(1-f^2)}=\omega_o\sqrt{1-f^2}.
$$    
For $\gamma>\gamma_c$, the minimal energy configuration in terms of variables $Q$ and $P$ reads  $(Q_{min},P_{min})=(\pm\sqrt{2(1-f^{-2})},0)$. Expanding the Hamiltonian around  this point gives
$$
H_{n'}=E_{min}^{n'}+\frac{1}{2}\frac{\omega_o(1+f^2)}{2j}P^2+\frac{1}{2}\omega_o 2j(f^2-1)Q^2+...,
$$   
therefore the pseudospin frequency is
$$
\omega_F(E_{min}^{n'})=\sqrt{\omega_o 2j (f^2-1)\frac{\omega_o(1+f^2)}{2j}}=\omega_o\sqrt{(f^2-1)(1+f^2)}=\omega_o\sqrt{f^4-1}.
$$
    
\section{Details of the semiclassical calculations of observables}
\label{appendixB}

In this appendix we present in detail  the calculations yielding to the  expressions for the expectation values of selected observables shown in section 4. 
\subsection{Fast pseudospin calculation}

First, we calculate the   denominator appearing in the expressions for the expectation values, which  is proportional to the  semiclassical approximation to    the Density of States [$\nu^f(E,m')$] or Weyl's formula
\begin{equation}
2\pi \nu^{f}(E,m')=\int\int dq\,dp\,\,\delta\left[E-H_{m'}(q,p)\right].
\label{eq:nua}
\end{equation}
The integration over $p$ can be performed, by expressing   the delta function in terms of  the zeros of the equation $E-H_{m'}(q,p)=0$, \begin{equation}
p_{\pm}=\pm \sqrt{\frac{2}{\omega}(E-V_{m'}(q))}.
\label{eq:zerosp}
\end{equation}
From where, we obtain
\begin{equation}
2\pi \nu^{f}(E,m')=\int\int dq\,dp\,\left(\frac{\delta\left(p-p_{+}\right)}{\left|\partial H_{m'}/\partial p\right|}_{p_{+}}+\frac{\delta\left(p-p_{-}\right)}{\left|\partial H_{m'}/\partial p\right|}_{p_{-}}\right).
\end{equation}
Since $\left|\partial H_{m'}/\partial p\right|_{p_{\pm}}=|\omega p_\pm|=\sqrt{2\omega(E-V_{m'}(q))}$, the integration over variable $p$ gives
\begin{equation}
2\pi\nu^{f}(E,m')=\sqrt{\frac{2}{\omega}}\int_{q\in \mathcal{I}_{E,m'}}\,\frac{dq}{\sqrt{E-V_{m'}(q)}},
\label{eq:nuar}
\end{equation}
provided that $p_\pm$ are real numbers. This latter  condition  is guaranteed by  the classically allowed integration region $q\in \mathcal{I}_{E,m'}$,   defined by  $E\geq V_{m'}(q)$. The boundaries of this integration region are obtained by solving 
$
E=V_{m'}(q)
$, which implies the following quadratic equation for the variable $q^2$
$$
E^2-\omega E q^2+\frac{\omega^2}{4}q^4=m'^2\left(\omega_o^2+\frac{4\gamma^2}{j}q^2\right),
$$
whose solutions are
$$
q^2_\pm=\left(\frac{2j\omega_o}{\omega}\right)\left(\frac{E}{\omega_o j}+\left(\frac{m'}{j}\right)^2 f^2\pm \frac{m'}{j}\sqrt{1+2\frac{E}{\omega_o j} f^2+\left(\frac{m'}{j}\right)^2 f^4}\right),
$$
where we have used the definition  $f=\gamma/\gamma_c$ with $\gamma_c=\sqrt{\omega\omega_o}/2$.  
For energies inside the double well potential, $-j\leq m'\leq -j/f^2$ and $E_{min}^{m'}<E\leq \omega_o m'$ [see panel (b) of Fig.\ref{fig:1}], the classically allowed region is formed by two disconnected intervals
$$
\mathcal{I}_{E,m'}=[-q_-(E,m'),-q_+(E,m')] \cup [q_+(E,m'),q_-(E,m')].
$$
For any other case the classical allowed region is given by the  interval 
$$
\mathcal{I}_{E,m'}=[-q_-(E,m'),q_-(E,m')].
$$

To calculate the expectation value of the bosonic  number operator, we substitute $a^\dagger a\rightarrow \frac{1}{2}(p^2+q^2)$ to obtain 
$$
 \langle a^\dagger a\rangle_{E,m'}=\frac{1}{2\pi \nu^{f}(E,m')}
\int dp dq \frac{1}{2}\left(p^2+q^2\right)\delta[E-H_{m'}(p,q)].
$$
As before,  we express the delta function in terms of the zeros of its argument (\ref{eq:zerosp}), to obtain
$$
\langle a^\dagger a\rangle_{E,m'}=\frac{1}{2\pi \nu^{f}(E,m')}\int dq dp \frac{1}{2}\left(p^2+q^2\right)\left(\frac{\delta\left(p-p_{+}\right)}{\left|\partial H_{m'}/\partial p\right|}_{p_{+}}+\frac{\delta\left(p-p_{-}\right)}{\left|\partial H_{m'}/\partial p\right|}_{p_{-}}\right).
$$
The integration over variable $p$ is straightforward 
$$
\int_{q\in\mathcal{I}_{E,m'}}\frac{1}{2}\left(\frac{p_+^2+q^2+p_-^2+q^2}{\sqrt{2\omega(E-V_{m'}(q))}}\right)dq=\frac{1}{2\sqrt{2\omega}}\int_{q\in\mathcal{I}_{E,m'}}\left(\frac{2 q^2+\frac{4}{\omega}(E-V_{m'}(q))}{\sqrt{E-V_{m'}(q)}}\right)dq,
$$ 
where we have used,  the interval defined above  $q\in \mathcal{I}_{E,m'}$ to guarantee that $p_\pm$ are real numbers. By using, in the numerator of the integrand,  the explicit expression of the potential $V_{m'}(q)$ (\ref{eq:vm}), we obtain after some direct simplifications the result 
$$\langle a^\dagger a\rangle_{E,m'}= \frac{1}{2\pi \nu^{f}(E,m')}\frac{1}{\omega}\sqrt{\frac{2}{\omega}}\int_{\mathcal{I}_{E,m'}} \frac{E-\sqrt{w_o^2+(\frac{2\gamma q}{\sqrt{j}})^2}}{\sqrt{E-V_{m'}(q)}}dq.
$$
To calculate the expectation value of $J_z$,  we take advantage of the inverse rotation $J_z=\cos\beta J_{z'}-\sin\beta J_{x'}$  to evaluate $$\langle j m'| J_z |j,m'\rangle=\cos\beta\  m'=\frac{\omega_o}{\sqrt{w_o^2+(\frac{2\gamma q}{\sqrt{j}})^2}}\ m',$$
where we have used $\langle j m'|J_{z'}|j m'\rangle=m'$, $\langle j m'|J_{x'}|j m'\rangle=0$, and the definition of  $\cos\beta$ in Eq.(\ref{eq:beta}).  With this  result the expectation value of $J_z$ for each $m'$ and given energy $E$ is
$$
\langle J_z \rangle_{E,m'}=\frac{1}{2\pi\nu^{f}(E,m')}
\int dp dq \frac{\omega_o \ m'}{\sqrt{w_o^2+(\frac{2\gamma q}{\sqrt{j}})^2}}\delta[E-H_{m'}(p,q)].
$$
The integral appearing in the previous expression
$$
\int  dq dp  \frac{\omega_o \ m'}{\sqrt{w_o^2+(\frac{2\gamma q}{\sqrt{j}})^2}}\delta[E-H_{m'}(p,q)]=\int dq \frac{\omega_o \ m'}{\sqrt{w_o^2+(\frac{2\gamma q}{\sqrt{j}})^2}}\int dp \delta[E-H_{m'}(p,q)], 
$$
has the  same   integral over the variable $p$ that  we have already calculated before (\ref{eq:nua}). Therefore, using the result of Eq.(\ref{eq:nuar}), we obtain  
$$
 \langle J_z \rangle_{E,m'}=\frac{1}{2\pi\nu^{f}(E,m')}\sqrt{\frac{2}{\omega}}\int_{\mathcal{I}_{E,m'}}\frac{\omega_o\ m' dq}{\sqrt{w_o^2+(\frac{2\gamma q}{\sqrt{j}})^2}\sqrt{E-V_{m'}(q)}}. 
$$ 
 
\subsection{Fast boson calculations}
As in the previous case, first we calculate the denominator appearing in the expression for the expectation values, which is proportional to the semiclassical approximation to the density of states $\nu^{a}(E,n')$ 
\begin{eqnarray}
2\pi \nu^{a}(E,n')&=\int d\phi\int_{-j}^j d j_z  \delta\left[E-H_{n'}\left(\vec{j}\right)\right]
 \\&=\int d\phi\int_{-j}^j d j_z  \left(\frac{\delta(j_z-j_{z+})}{\left|\partial H_{n'}/\partial j_z\right|_{j_{z+}}}+\frac{\delta(j_z-j_{z-})}{\left|\partial H_{n'}/\partial j_z\right|_{j_{z-}}}\right),
\end{eqnarray}
where $j_{z\pm}$ are the solutions of the equation $E-H_{n'}(j_z,\phi)=0$,   given by
$$
\frac{j_{z\pm}}{j} f^2\cos^2\phi \ +1 =\pm
\sqrt{\mathcal{F}(\cos\theta)}
$$
where we have used the  definition  $\mathcal{F}(x)\equiv 1+2f^{2}\epsilon_{n'} x^2+f^{4}x^{4}$,
with   $\epsilon_{n'}\equiv (E-\omega n')/(\omega_o j)$. 
Calculating the derivatives
$$
\partial H_{n'}/\partial j_z=\omega_o\left( \frac{j_z}{j}f^2\cos^2\phi+1\right),
$$
and evaluating them in $j_{z\pm}$, we obtain
$$
\left|\partial H_{n'}/\partial j_z\right|_{j_{z\pm}}=\omega_o\sqrt{1+2\cos^2\phi \epsilon_{n'}f^2+\cos^4\phi f^4}\equiv\omega_o \sqrt{\mathcal{F}(\cos\phi)}.
$$
 Therefore the integral reads
$$
2\pi \nu^{a}(\epsilon,n')=\frac{1}{\omega_o}\int d\phi\int_{-j}^j d j_z \frac{\delta(j_z-j_{z+})+\delta(j_z-j_{z-})}{\sqrt{\mathcal{F}(\cos\phi)}}. 
$$
To perform the integration over the variable $j_z$, we have to investigate when the solutions $j_{z\pm}$ are in the interval $j_z\in[-j,j]$. We can distinguish two cases depending on the value of $\epsilon_{n'}$. If $\epsilon_{n'}<-1$, the two solution $j_{z\pm}$ are in the interval $[-j,j]$ if and only if  the square root argument appearing in the denominator is greater or equal  than 0: $\mathcal{F}(\cos\phi)=1+2\cos^2\phi\, \epsilon_{n'} f^2+\cos^4\phi f^4\geq 0$.  The previous condition is satisfied in two intervals around $\phi=0$ and $\phi=\pi$
$$
\phi\in [-\phi_o,\phi_o]\cup [\pi-\phi_o,\pi+\phi_o],
$$ 
where $\phi_o$ is  the solution of the equation $1+2\cos^2\phi\, \epsilon_{n'} f^2+\cos^4\phi f^4=0$, given by
$$
\cos^{2}(\phi_o)=\frac{1}{f^2}\left[-\epsilon_{n'}+\sqrt{\epsilon_{n'}^{2}-1}\,\right].
$$
Therefore in this case, $\epsilon_{n'}<-1$, the integral over $j_z$ gives
$$
2\pi \nu^{a}(E,n')=\frac{1}{\omega_o}\int_{\phi\in [-\phi_o,\phi_o]\cup [\pi-\phi_o,\pi+\phi_o]} d\phi \frac{2}{\sqrt{\mathcal{F}(\cos\phi)}}. 
$$
Now, since the dependence on $\phi$ in the integrand enters through $\cos^2\phi$, the integral over $\phi\in [-\phi_o,\phi_o]\cup [\pi-\phi_o,\pi+\phi_o]$, can be expressed as four times the integral in the interval $\phi\in[0,\phi_o]$. In this way the final form for the integral is
$$
2\pi \nu^{a}(E,n')=\frac{8}{\omega_o}\int_{0}^{\phi_o}  \frac{d\phi}{\sqrt{\mathcal{F}(\cos\phi)}}. 
$$
For $-1<\epsilon_{n'}\leq 1$, only the solution $j_{z+}$ is in the interval $[-j,-j]$, and that for any value of $\phi\in[0,2\pi]$. Therefore in this case the integral takes the form
$$
2\pi \nu^{a}(E,n')=\frac{1}{\omega_o}\int_{0}^{2\pi} d\phi \frac{1}{\sqrt{\mathcal{F}(\cos\phi)}}. 
$$
Gathering the previous results, we obtain the Eq.(\ref{eq:denFB}). 

The  expectation value of $J_{z}$, 
$$
\langle J_z\rangle_{n',E}=\frac{1}{2\pi\nu^a(E,n')}\int dj_zd\phi \ j_z\delta\left[E-H_{n'}\left(\vec{j}\right)\right],
$$
can be calculated similarly. By expressing the Dirac delta in terms of the roots of its argument, $j_{z\pm}$, we obtain
$$
\frac{1}{2\pi\nu^a(E,n')}\displaystyle\int d\phi \displaystyle\int_{-j}^j dj_z  \frac{\delta(j_z-j_{z+})+\delta(j_z-j_{z-})}{\omega_o\sqrt{\mathcal{F}(\cos\phi)}}j_z.
$$
As before, in the case $\epsilon_{n'}\leq -1$ the two roots are in $[-j,j]$ for $\phi$ in the region defined above, therefore the integral gives
$$
\frac{4}{2\pi\nu^a(E,n')\omega_o }\displaystyle\int_0^{\phi_o} d\phi \frac{j_{z+}+j_{z-}}{\sqrt{\mathcal{F}(\cos\phi)}}.
$$
The sum of the roots is $j_{z+}+j_{z-}=-2 j/(f^2 \cos^2\phi)$, yielding to the expression  
$$
\langle J_z\rangle_{n',E}=-\frac{8j}{f^2 \omega_o 2\pi\nu^a(E,n')}\displaystyle\int_0^{\phi_o}  \frac{d\phi}{\cos^2\phi\sqrt{\mathcal{F}(\cos\phi)}}.
$$
For $-1<\epsilon_{n'}\leq 1$, since only the root $j_{z+}$ is in the interval $[-j,j]$ and $\phi$ is not restricted, the expectation value is
$$
\langle J_z\rangle_{n',E}=\frac{1}{2\pi\nu^a(E,n')\omega_o }\displaystyle\int_0^{2\pi} d\phi \frac{j_{z+}}{\sqrt{\mathcal{F}(\cos\phi)}}=-\frac{j}{f^{2}\omega_{0}2\pi\nu^{a}(E,n')}\displaystyle\int_0^{2\pi}d\phi\frac{ 1-\sqrt{\mathcal{F}(\cos\phi)}}{\cos^2(\phi)\sqrt{\mathcal{F}(\cos\phi)}}.
$$
The  summary of these results is shown in Eq.(\ref{eq:jzFB}).

To calculate the expectation value of  the boson number operator, $a^\dagger a$,  we write it in terms of the $b^{\dagger}$ and $b$ operators
\begin{equation}
a^\dagger a=b^\dagger b+\frac{2\gamma^2}{j\omega}j_x^2-\sqrt{\frac{2}{j}}\frac{\gamma}{\omega}j_x (b^\dagger +b),
\end{equation}

to calculate $\langle n'| a^\dagger a|n'\rangle=n'+\frac{2\gamma^2}{j\omega}j_x^2$, where we have used that $\langle n'|(b^\dagger+b)|n'\rangle=0$ and $\langle n'|b^\dagger b|n'\rangle=n'$.  With the previous result,  the expectation value of the   number of photons in the fast boson approximation reads
\begin{eqnarray}
&
\langle a^\dagger a\rangle_{E,n'}=
\frac{1}{2\pi\nu^{a}(E,n')}
\int d j_z d\phi \left( n'+ \frac{2\gamma^2}{j\omega}j_x^2\right) \delta \left[E-H_{n'}\left(\vec{j}\right)\right]\nonumber\\
&=n'+\frac{2\gamma^2}{ 2\pi \nu^{a}(E,n')j\omega }\displaystyle\int dj_zd \phi (j^2-j_z^2)\cos^2\phi \delta\left[(E-H_{n'}\left(\vec{j}\right)\right],
\end{eqnarray}
where we have expressed the $j_x$ component in terms of the canonical variables $j_x=\sqrt{j^2-j_z^2}\cos\phi$.

The two cases previously identified have to be considered separately to evaluate the integral of the second term. In the case $\epsilon_{n'}\leq 1$ the integral is
\begin{eqnarray}
&\frac{2\gamma^2}{ 2\pi \nu^{a}(E,n')j\omega }\displaystyle\int dj_zd \phi (j^2-j_z^2)\cos^2\phi \delta\left[(E-H_{n'}\left(\vec{j}\right)\right]\nonumber\\
&=
\frac{2\gamma^2}{ 2\pi \nu^{a}(E,n')j \omega }4\int_{0}^{\phi_o}\frac{j^2-j_{z+}^2+j^2-j_{z-}^2}{\omega_o\sqrt{\mathcal{F}(\cos\phi)}}\cos^2\phi d\phi.\nonumber
\end{eqnarray}
Now, by using the expressions for the roots $j_{z\pm}$ and the definition of $\mathcal{F}(\cos\phi)$, we obtain $$j^2-j_{z+}^2+j^2-j_{z-}^2=-4 j^2\frac{1+f^2\epsilon_{n'}\cos^2\phi}{f^4\cos^4\phi}.
$$
By substituting this result in the integral, we obtain
$$
-\frac{2\gamma^2}{ 2\pi \nu^{a}(E,n')j \omega }\left(\frac{16 j^2}{f^4\omega_o}\right)\int_{0}^{\phi_o}\frac{1+f^2\epsilon_{n'}\cos^2\phi}{\cos^2\phi\sqrt{\mathcal{F}(\cos\phi)}} d\phi.\nonumber
$$
Recalling that $f=\gamma/\gamma_c$ with $\gamma_c=\sqrt{\omega\omega_o}/2$, we   simplify the factor in front of the previous integral to obtain  the following expression for the expectation value of $a^\dagger a $, valid  in the case $\epsilon_{n'}\leq 1$
$$ 
\langle a^\dagger a\rangle_{E,n'}=n'-\frac{8j}{f^2 2\pi\nu^{a}(E,n')}\displaystyle\int_0^{\phi_o}d\phi \frac{1+f^2\epsilon_{n'} \cos^2\phi }{\cos^2\phi \sqrt{\mathcal{F}(\cos\phi)}}.
$$
For $-1<\epsilon_{n'}\leq 1$, the integral is
\begin{eqnarray}
&\frac{2\gamma^2}{ 2\pi \nu^{a}(E,n')j\omega }\displaystyle\int dj_zd \phi (j^2-j_z^2)\cos^2\phi \delta\left[(E-H_{n'}\left(\vec{j}\right)\right]\nonumber\\
&=
\frac{2\gamma^2}{ 2\pi \nu^{a}(E,n')j \omega }\int_{0}^{2\pi}d\phi\frac{j^2-j_{z+}^2}{\omega_o\sqrt{\mathcal{F}(\cos\phi)}}\cos^2\phi .\nonumber
\end{eqnarray}
By using the expression for the root $j_{z+}$ and the definitions of $\mathcal{F}$ and $f$, the previous integral simplifies to yield the following expression
$$ 
\langle a^\dagger a\rangle_{E,n'}=n'-
\frac{j}{f^2 2\pi\nu^{a}(E,n')}\displaystyle\int_0^{2\pi}d\phi \frac{1+\epsilon_{n'} f^2\cos^2\phi-\sqrt{\mathcal{F}(\cos\phi)}}{\cos^2\phi\sqrt{\mathcal{F}(\cos\phi)}}.
$$
The previous results are gathered in Eq(.\ref{eq:aaFB}).

\section{Classical equations of motion of the Dicke model}
\label{appendixC}
The full classical Dicke Hamiltonian that we use in this contribution is the one obtained from a semiclassical approximation to the quantum propagator written in terms of coherent states \cite{Ribeiro06}
$$
H_{cl}=\langle\alpha;z| H_D | \alpha,z \rangle,
$$ 
where $|\alpha;z\rangle=|\alpha\rangle\otimes|z\rangle$ with $|z\rangle$ and $|\alpha\rangle$ Bloch (for the pseudospin) and Glauber (for the bosons) coherent states respectively
\begin{eqnarray}
|z\rangle&=&\frac{1}{(1+|z|^2)^j}e^{z J_+}|j,-j\rangle \nonumber\\
|\alpha\rangle&=&e^{-|\alpha|^2/2}e^{\alpha a^\dagger} |0\rangle.
\end{eqnarray}
The expectation value of the Dicke Hamiltonian in the coherent states reads (we take $\mathcal{N}=2j$)
$$
H_{cl}= \omega |\alpha|^2-\omega_{o}j \left(\frac{1-|z|^2}{1+|z^2|} \right)+\frac{2\gamma}{\sqrt{2j}}(\alpha+\alpha^*)j\frac{z+z^*}{1+|z^2|}.
$$  
We consider   canonical variables $(q,p)$ and $(\phi,j_z)$, related to the complex coherent parameters through
\begin{eqnarray}
\alpha &=&\frac{1}{\sqrt{2}}(q+ip) \nonumber \\
z&=&\tan(\theta/2)e^{-i\phi}=\sqrt{\frac{1+(j_z/j)}{1-(j_z/j)}}e^{-i\phi}.
\end{eqnarray}
The expectation value of the Hamiltonian, written in terms of the previous variables define the classical Hamiltonian
$$
H_{cl}(q,p,\phi,j_z)=\frac{\omega}{2}(p^2+q^2)+\omega_o j_z+\frac{2\gamma}{\sqrt{j}}q\sqrt{j^2-j_z^2}\cos\phi,
$$
with classical  Hamilton equations given by
$$
dq/dt=\partial_p H_{cl}\ \ \ \ dp/dt=-\partial_q H_{cl}
$$
$$
d\phi/dt=\partial_{j_z} H_{cl} \ \ \ dj_z/dt=-\partial_\phi H_{cl}.
$$
  

\end{document}